\documentclass[aps,prd,superscriptaddress,preprintnumbers,nofootinbib,10pt]{revtex4-2}
\usepackage{multirow}
\usepackage{amsmath}
\usepackage{amssymb}
\usepackage[dvipdf,dvips]{graphicx}
\usepackage{color}
\usepackage{hyperref}
\usepackage{url}
\usepackage{slashed}
\usepackage[usenames,dvipsnames]{xcolor}
\usepackage{amsmath}
\usepackage{amsfonts}
\usepackage{float} 
\usepackage{amssymb}
\usepackage{epsfig}
\usepackage{graphics}
\usepackage{euscript}
\usepackage{slashed}
\usepackage{epstopdf}
\usepackage[utf8]{inputenc}
\allowdisplaybreaks
\usepackage[normalem]{ulem}
\usepackage{pifont}
\usepackage{dsfont}
\usepackage{MnSymbol}
\usepackage{graphicx}
\usepackage{subcaption}
\usepackage{latexsym}
\usepackage{tikz-feynman}
\usepackage{tikz-cd}
\usepackage{easyReview}
\usepackage{cancel}
\usepackage[normalem]{ulem}
\usepackage{svg}
\usepackage{cleveref}

\newcommand{\LambdaMS}{\Lambda_{\overline{\text{MS}}}}

\makeatletter
\DeclareRobustCommand{\cev}[1]{%
	{\mathpalette\do@cev{#1}}%
}
\newcommand{\do@cev}[2]{%
	\vbox{\offinterlineskip
		\sbox\z@{$\m@th#1 x$}%
		\ialign{##\cr
			\hidewidth\reflectbox{$\m@th#1\vec{}\mkern4mu$}\hidewidth\cr
			\noalign{\kern-\ht\z@}
			$\m@th#1#2$\cr
		}%
	}%
}
\makeatother

\makeatletter
\newlength{\negph@wd}
\DeclareRobustCommand{\negphantom}[1]{%
  \ifmmode
    \mathpalette\negph@math{#1}%
  \else
    \negph@do{#1}%
  \fi
}
\newcommand{\negph@math}[2]{\negph@do{$\m@th#1#2$}}
\newcommand{\negph@do}[1]{%
  \settowidth{\negph@wd}{#1}%
  \hspace*{-\negph@wd}%
}
\makeatother

\newcommand{\msbar}{{\overline{\mbox{{MS}}}}}
\newcommand{\lms}{\Lambda_{\overline{\mbox{\tiny{MS}}}}}

\begin{document}

\title{Casimir effect in Gribov-Zwanziger theory}


\author{David Dudal}
\email{david.dudal@kuleuven.be} 
\affiliation{KU Leuven Campus Kortrijk-Kulak, Department of Physics,	Etienne Sabbelaan 53 bus 7657, 8500 Kortrijk, Belgium}

\author{Philipe De Fabritiis} 
\email{pdf321@cbpf.br} 
\affiliation{CBPF $-$ Centro Brasileiro de Pesquisas Físicas, Rua Dr. Xavier Sigaud 150, 22290-180, Rio de Janeiro, Brazil}
\affiliation{KU Leuven Campus Kortrijk-Kulak, Department of Physics,	Etienne Sabbelaan 53 bus 7657, 8500 Kortrijk, Belgium}
	
\author{Sebbe Stouten}
\email{sebbe.stouten@kuleuven.be} 
\affiliation{KU Leuven Campus Kortrijk-Kulak, Department of Physics,	Etienne Sabbelaan 53 bus 7657, 8500 Kortrijk, Belgium}

\author{David Vercauteren}
\email{vercauterendavid@duytan.edu.vn}
\affiliation{Institute of Research and Development, Duy Tan University, Da Nang 550000, Vietnam}
\affiliation{Faculty of Natural Sciences, Duy Tan University, Da Nang 550000, Vietnam}

\begin{abstract}
We consider Yang-Mills theory with two infinite parallel plates, separated by a distance \(L\), that are perfect magnetic conductors (PMC) or perfect electric conductors (PEC). Recently, it was shown that the Gribov copy problem persists in such a setting.   We then study the Gribov-Zwanziger (GZ) action in the presence of those boundaries using functional integral methods. Lagrange multiplier fields allow one to lift the boundary conditions into the action, after which the boundary modifications to the gluon propagator can straightforwardly be determined. In the PEC case, we provide evidence that, even when translation invariance is (partially) broken, the usual horizon term in the GZ action still restricts the functional integral to the Gribov region. We compute the Casimir energy for GZ with PMC or PEC plates, both directly from the functional integral and from the energy-momentum tensor, obtaining consistent results. We compare our analytical results with recent lattice data, in both 4D and 3D. A priori, one might expect that the boundary-modified gluon propagator introduces new \(L\)-dependencies into the GZ gap equation. This would make the Gribov mass \(\gamma\) dynamically dependent on \(L\), implying an interesting interplay with the Casimir energy. However, we show that no such dynamical \(L\)-dependence occurs within the current approximation. 
\end{abstract}

\maketitle	

\section{Introduction}	

Quantum Chromodynamics (QCD) is the quantum field theory (QFT) that describes the strong interactions between quarks and gluons~\cite{Gross2023}. Despite its tremendous success over the past decades in the comparison between theoretical predictions and experiments, QCD harbors at its core one of the greatest mysteries of theoretical physics: the confinement problem~\cite{Greensite11}. Schematically, the hadrons that we observe in the laboratory are described in terms of quarks and gluons, degrees of freedom that were never directly observed, and the underlying mechanism that confines these elusive particles inside the abundant hadrons is still poorly understood.

Although this challenging problem is still awaiting a satisfactory explanation from first principles, there are interesting numerical findings that might give some clues. For instance, lattice simulations for pure Yang-Mills (YM) theories showed that the potential between static heavy quarks increases linearly with the distance between them, making the formation of bound states energetically favorable~\cite{Greensite:2003bk}. A physically appealing scenario to explain this linear confining potential is the formation of a chromoelectric flux-tube between the quarks~\cite{Mandelstam:1974pi,Nambu74,tHooft:1981bkw}, which closely resembles the magnetic flux tubes that usually emerge in superconductors; this is why such a conjectured mechanism for confinement is typically called the dual superconductor picture of the QCD vacuum~\cite{Baker:1991bc}.

A simple yet insightful phenomenological framework to describe the QCD vacuum is provided by the MIT bag model~\cite{PhysRevD.9.3471}. In this approach, unobservable colored degrees of freedom (quarks and gluons) are confined within an abstract bag, which serves as a representation of the observable colorless hadronic bound states. While this framework offers an appealing and intuitive way of capturing the essence of confinement, it immediately raises a fundamental question: what physical mechanism ensures that these colored particles remain bound and that the bag itself is stable? One intriguing possibility is the Casimir effect~\cite{Casimir:1948dh}, a subtle and exciting consequence of quantum vacuum fluctuations.

The Casimir effect is a striking embodiment of the non-emptiness of the quantum vacuum. In 1948, Casimir showed that quantum fluctuations of the electromagnetic field can give rise to a macroscopic manifestation: two electrically neutral and perfectly conducting plates, placed close to each other, experience an attractive force between them~\cite{Casimir:1948dh}. This force emerges from the difference in the zero-point energy of the electromagnetic field with and without the presence of boundaries~\cite{Casimir:1948dh}. Although theoretically compelling, it took several decades before an experimental confirmation was achieved. Indeed, the first clear measurement came in 1997~\cite{Lamoreaux1997}, marking a milestone for the subject. Since then, experimental techniques have advanced considerably, leading to increasingly precise observations~\cite{Bimonte:2021sib}.

A novel and versatile framework for computing the Casimir energy through functional methods was recently developed in~\cite{Dudal:2020yah}, drawing inspiration from the pioneering work~\cite{Bordag:1983zk}, see also \cite{Golestanian_1998}. Within the functional integral formalism, boundary conditions can be incorporated directly into the action without breaking gauge invariance, by introducing auxiliary fields that serve as Lagrange multipliers. Once the gauge field is integrated out, this procedure yields a non-local effective boundary action from which the Casimir energy can be computed in a straightforward manner. The boundary dynamics approach described above has already proven successful in a variety of contexts, see e.g.~\cite{Canfora:2022xcx, Oosthuyse:2023mbs, Dudal:2024PEMC, Dudal:2024Robin, Dudal:2024DEM, Canfora:2025ssr, Dudal2026, DeFabritiis26}.

The idea of accounting for the stability of the MIT bag model through the Casimir effect has roots in earlier work~\cite{Plunien:1986ca}. Indeed, a standard perturbative computation of the YM Casimir effect for a spherical bag was already carried out before~\cite{Boyer68,Davies72,Bender76}, yielding a repulsive force that is clearly incompatible with the looked-for stability of the bag. However, in later research~\cite{Oxman05, Canfora13} it was shown that an attractive Casimir force can be obtained by incorporating non-perturbative modifications to the gluon propagator, highlighting the crucial role of infrared non-perturbative dynamics in the non-Abelian Casimir effect. Furthermore, recent lattice studies on the non-Abelian Casimir effect~\cite{Chernodub:2018pmt, Chernodub:2023dok,Ngwenya:2025cuw,Ngwenya:2025mpo, Chernodub:2019nct} reveal the emergence of a new intrinsic mass scale, pointing to novel non-perturbative features that still await a complete analytical understanding. Therefore, this motivates us to provide a fresh look at the non-Abelian Casimir effect by applying the novel functional techniques developed in Refs.~\cite{Dudal:2020yah, Canfora:2022xcx, Oosthuyse:2023mbs, Dudal:2024PEMC, Dudal:2024Robin, Dudal:2024DEM, Canfora:2025ssr, Dudal2026} within models tailored to capture the infrared dynamics of YM theories.

The quantization of YM theories through the functional integral formalism is traditionally carried out using the Faddeev-Popov (FP) gauge fixing procedure. While highly successful in the perturbative domain, the assumptions underlying the FP quantization are not well-grounded in the non-perturbative realm. Indeed, the infrared regime is plagued by the Gribov problem~\cite{Gribov78}, namely the impossibility of selecting a unique representative of a gauge orbit in covariant gauges~\cite{Singer78}, which renders the FP approach ill-defined. A first-principle solution to deal with such redundancies was proposed at leading order by Gribov in 1978~\cite{Gribov78} and later extended to all orders by Zwanziger in 1989~\cite{Zwanziger89}. The resulting Gribov-Zwanziger (GZ) framework~\cite{Vandersickel:2012tz} provides a local and renormalizable action that improves FP quantization by explicitly accounting for the infinitesimal Gribov copies that jeopardize the FP procedure in the infrared regime. In this work, we adopt the GZ framework, but it is worthy pointing out that several other approaches have been developed to deal with the infrared dynamics of YM theories, including Lattice QCD~\cite{Cucchieri:2007rg,Boucaud:2011ug}, Schwinger-Dyson equations~\cite{Fischer:2006ub}, the Functional Renormalization Group~\cite{Fischer:2008uz}, massive versions of Yang-Mills \cite{Comitini:2020ozt}, and the Curci-Ferrari model~\cite{Pelaez:2021tpq, Reinosa:2024vph}, among others.

In recent work \cite{Dudal:2026uuv}, the persistence of the Gribov copy problem in presence of parallel plates was discussed. The main goal of the current work is to provide a first look into non-perturbative features of the non-Abelian Casimir effect in relation to the Gribov copy problem. To achieve this, we employ the GZ framework as well as functional integral methods that are suitable to explore the effective boundary dynamics in the presence of non-trivial boundaries, devoting particular attention to the role played by the Gribov mass.

This paper is organized as follows. In Sec.~\ref{sec:GZ} we investigate the interplay between the Gribov-Zwanziger action and boundary conditions. More specifically, in Sec.~\ref{sec:GZ+BC}, we briefly review the GZ framework in 4D and assume at first the underlying no-pole condition does not change upon adding planar boundary conditions. We then introduce auxiliary fields to enforce the standard perfect electric conductor (PEC) or perfect magnetic conductor (PMC) boundary conditions. Integrating out all fields except the auxiliary ones, we retrieve an 3D effective boundary action in Sec.~\ref{sec:3d-bnd-action}. The modifications in the gluon propagator due to the boundary conditions are discussed in Sec.~\ref{sec:A-prop}.  Assuming that the Gribov mass does not depend on the plate distance,  in Sec.~\ref{sec:free-Gribov-param}, we compute the Casimir energy for both PEC and PMC plates, using two different methods: directly from the functional integral~\ref{sec:Casimir-direct}, and via the energy-momentum tensor~\ref{sec:Casimir-EMT}. Then, we argue in Sec.~\ref{sec:solving-gap-eq} that the gap equation is not modified by the presence of the boundary, implying that the Gribov mass is indeed independent of the plate distance. We also solve the gap equation explicitly in the so-called $V$-scheme, a physical renormalization scheme. We compare our findings with recent lattice data and also briefly analyze the 3D case. In Sec.~\ref{sec:no-pole-condition}, we investigate in more detail the effect of the boundary condition on the no-pole condition for the PEC case, confirming at leading order our earlier assumption made sense. In Sec.~\ref{sec:conclusion}, we report our conclusions, and discuss some interesting directions for future research.

\section{General framework}\label{sec:GZ}

The main conventions adopted in this work are the following. Four-vectors are written in normal font \(k=(k_t,k_x,k_y,k_z)\) and three-vectors in bold \(\mathbf{k} = (k_t,k_x,k_y)\). Greek indices run over \(\{t,x,y,z\}\) and Latin indices over \(\{t,x,y\}\). Superscripts \(a,b,c,\dots\) indicate \(SU(N)\) color indices \(\{1,\dots,N^2-1\}\), and superscripts \(\gamma,\lambda\) denote plate indices \(\{-,+\}\).
For the Fourier transformation, we write
\begin{equation}
	X_i(x) = \int \frac{d^d{k}}{(2\pi)^d} \hat{X}_i(k)e^{-ik\cdot x}
\end{equation}
for a \(d\)-dimensional vector field \(X\), but we will drop the hat in order to not overload notations. In this convention, the Dirac delta becomes
\begin{equation}
	\delta(z) = \int \frac{d{k_z}}{2\pi} e^{-ik_z z}.
\end{equation}

\subsection{Gribov-Zwanziger action with a boundary term}\label{sec:GZ+BC}
\subsubsection{The Gribov-Zwanziger framework (without boundaries)}

Let us consider the classical $SU(N)$ YM action in a $d$-dimensional Euclidean spacetime,
\begin{align}
    S_\text{YM} = \int \!\! d^dx \, \frac{1}{4} F_{\mu \nu}^a F_{\mu \nu}^a,
\end{align}
where $F_{\mu \nu}^a = \partial_\mu A_\nu^a - \partial_\nu A_\mu^a + g f^{abc} A_\mu^b A_\nu^c$ is the non-Abelian field strength tensor. Quantizing $ S_\text{YM}$ through the functional integral formalism, following the FP method for the gauge-fixing adopting the linear covariant gauge, we find
\begin{align}
    S_\text{FP} = S_\text{YM} + \int \! d^dx \left(- \frac{\alpha}{2} \kappa^a \kappa^a + \kappa^a \partial_\mu A_\mu^a + \bar{c}^a \partial_\mu D_\mu^{ab} c^b\right),
\end{align}
where $D_\mu^{ab} = \delta^{ab} \partial_\mu - g f^{abc}A_\mu^c$ is the covariant derivative in the adjoint representation, $\left(c^a, \bar{c}^a\right)$ are the FP ghosts, $\alpha$ is a gauge parameter (throughout this paper, the Landau gauge limit $\alpha \rightarrow 0$ is always implicitly understood), and  $\kappa^a$ is the Nakanishi-Lautrup field, whose role is to enforce the gauge-fixing condition. Thus, the functional integral governing YM theory is
\begin{align}
    Z_\text{FP} = \int \! \mathcal{D}A \, \mathcal{D}c \, \mathcal{D}\bar{c} \, \mathcal{D}\kappa \, e^{-S_\text{FP}}.
\end{align}

The FP action boasts excellent predictive power in the ultraviolet regime, where, thanks to asymptotic freedom, perturbation theory works well. However, in the infrared regime, the story is more complicated. Indeed, in 1978, Gribov showed that in the Landau gauge, the FP procedure does not completely fix the gauge~\cite{Gribov78}. Afterwards, Singer demonstrated that, rather than being a peculiarity of the Landau gauge, this issue will be present for a more general class of gauge-fixing conditions~\cite{Singer78}. This means that there are different field configurations, all satisfying the same gauge condition, that are nevertheless connected by gauge transformations: the so-called Gribov copies. In order to keep this paper self-contained, we will give a brief discussion of how to handle the existence of Gribov copies. A more detailed treatment can be found, for example, in Refs.~\cite{Sobreiro05,Vandersickel11}.

For infinitesimal gauge transformations, the Gribov copies are directly related to zero-modes of the Landau gauge FP operator, $\mathcal{M}^{ab} = - \partial_\mu D_\mu^{ab}$. This hampers the FP procedure itself, since the FP operator is assumed to be a positive operator in order to serve as a measure in the functional integral. A possible way to deal with these (infinitesimal) copies is to restrict the functional integral to the so-called Gribov region:
\begin{align}
    \Omega = \{A_\mu^a \big\vert \, \partial_\mu A_\mu^a = 0 \,\,{\rm and}\,\, \mathcal{M}^{ab} > 0\}.
\end{align}

Gribov himself performed this restriction at first order in perturbation theory using the famous no-pole condition for the ghost propagator $G^{ab}(p^2)$~\cite{Gribov78}. Roughly speaking, the latter is the inverse of the FP operator. Upon using the common parameterization,
\begin{align}
G^{ab}(p^2)=\frac{\delta^{ab}}{p^2\left[1-\sigma(p^2)\right]},
\end{align}
the no-pole condition then reads $\sigma(p^2)<1$ for $p^2>0$.

A decade later, Zwanziger generalized this result to all orders in perturbation theory following another strategy~\cite{Zwanziger89}. Interestingly enough, it was shown later that Gribov's strategy can be generalized to all orders, leading to the same result Zwanziger obtained, being a highly non-trivial check of its validity \cite{Capri13}.

In summary, the restriction to the Gribov region can be achieved by the introduction of a non-local term
\begin{align}\label{eq:action-GZ-non-local}
    S_\text{GZ}^\text{NL} = S_\text{FP} + \gamma^4 H(A) - \gamma^4 d V \left(N^2-1\right),
\end{align}
where $V$ is the $d$-dimensional spacetime volume and  the non-local horizon function \(H(A)\) is given by
\begin{align}\label{eq:horizon-function}
     H(A) \!=\!\! \int \!\! d^dx \, d^dy \, g f^{abc} A_\mu^b(x) \left[\mathcal{M}^{-1}_{xy}\right]^{a d} \! g f^{d e c} A_\mu^e(y).
\end{align}
There is a massive parameter $\gamma$, the so-called Gribov mass, emerging from the restriction to the Gribov region. The Gribov mass is not a free parameter; it must be fixed self-consistently through the gap equation
\begin{align}\label{eq:gap-eq-horizon}
    \langle H(A) \rangle = d V (N^2 - 1),
\end{align}
 where the vacuum expectation value (VEV) must be evaluated with the measure defined by the action in Eq.~\eqref{eq:action-GZ-non-local}. The gap equation can be rewritten in a more suggestive form:
\begin{align}\label{VacuumGapEquation}
    \frac{\partial \mathcal{E}}{\partial \gamma^2} = 0,
\end{align}
with the vacuum energy density \(\mathcal{E}\) defined by
\begin{align}
	e^{-V \mathcal{E}}  = \int \! \mathcal{D}\!A \, \mathcal{D}c \, \mathcal{D}\bar{c} \, \mathcal{D}\kappa \,\, e^{-S_\text{GZ}^\text{NL}}.
\end{align}
Therefore, the gap equation imposes that \(\gamma\) dynamically extremizes the vacuum energy density of the theory. This GZ procedure effectively implements the no-pole condition at the level of the path integral. The usual perturbative FP action is recovered in the $\gamma\to0$ limit.

The appearance of the inverse FP operator in the horizon function~\eqref{eq:horizon-function} renders the GZ action non-local, an inconvenient feature when using standard QFT tools. However, it has been shown that the GZ action can be rewritten in a local form by introducing auxiliary bosonic fields \((\varphi_\mu^{ab},\bar \varphi_\mu^{ab})\) and Grassmannian fields \((\omega_\mu^{ab},\bar \omega_\mu^{ab})\). That is, one can rewrite the non-local GZ action~\eqref{eq:action-GZ-non-local} in $d$ dimensions as 
\begin{align}\label{eq:action-GZ}
        S_\text{GZ} = S_\text{FP} + S_\text{H} - \gamma^4 d V \left(N^2-1\right),
\end{align}
where the horizon action, recast in a local form, is now given by the expression 
\begin{align}\label{eq:action-horizon-local}
    S_\text{H} \!=\! \int \! d^dx \, \left[-\bar{\varphi}^{ac}_\mu \mathcal{M}^{ab}\varphi^{bc}_\mu + \bar{\omega}^{ac}_\mu \mathcal{M}^{ab} \omega^{bc}_\mu  + \gamma^2 g f^{abc} A_\mu^a (\varphi + \bar{\varphi})_\mu^{bc}\right].
\end{align}
Therefore, the GZ partition function for YM theory that takes into account the presence of infinitesimal Gribov copies in a local and renormalizable action can be written as
\begin{align}
     Z_\text{GZ} = \!\! \int \!\! \mathcal{D}\Phi \, e^{-S_\text{GZ}},
\end{align}
where $S_\text{GZ}$ is given by Eq.~\eqref{eq:action-GZ} and the functional integral is performed over all fields, including the auxiliary ones, with the Gribov mass $\gamma$ being fixed through the gap equation~\eqref{VacuumGapEquation}. 

\begin{figure}
    \centering
    \includegraphics[width=0.35\linewidth]{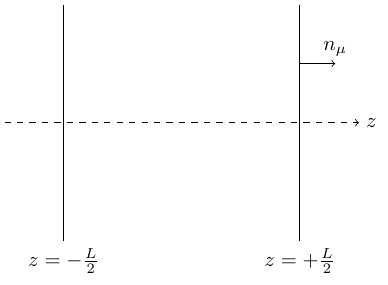}
    \caption{Illustration of the boundary: two infinite parallel plates at \(z=\pm \frac{L}2\), with unit normal vector \(n_\mu = \delta_{\mu z}\).}
    \label{fig:set-up-plates}
\end{figure}

\subsubsection{Introducing the boundary in 4D}

The next step is to introduce boundary conditions into the GZ action. It is important to stress that this comes with an assumption: the GZ action and associated gap equation effectively implement the no-pole condition. The derivation thereof was under the assumption of full translational invariance. Of course, adding boundaries affects the latter and this might also change the implications of the no-pole condition. So for now, we maintain the GZ action as is, and we will come back to the no-pole condition in the presence of planar boundaries in Sec.~\ref{sec:no-pole-condition}.
 
We thus consider two infinitely thin and long parallel plates located at \(z=\pm \frac{L}2\), with unit normal vector \(n_\mu = \delta_{\mu z}\), see Fig.~\ref{fig:set-up-plates}. On these plates, we will impose either perfect electric conductor (PEC) or perfect magnetic conductor (PMC)  boundary conditions\footnote{The reason we study both PMC and PEC is that the GZ action is not duality-invariant, not even at leading order, which means that they might exhibit different behaviors for each case (unlike in QED, which has a duality-invariant action). See, for instance,~\cite{Dudal2026,deser1976duality}.}, respectively defined by
\begin{align}\label{eq:PEC-PMC}
    \begin{split}
        \text{PEC}: \quad &\widetilde{F}^a_{\mu\nu}n_\nu = 0, \\*
	   \text{PMC}: \quad &F_{\mu\nu}^an_\nu = 0,
    \end{split}
\end{align}
in which the dual field strength is given by\footnote{The factor \(i\) shows up because we have Wick-rotated the Levi-Civita symbol.} $\widetilde{F}^a_{\mu\nu} = \frac{i}{2} \varepsilon_{\mu\nu\alpha\beta} F^a_{\alpha\beta}$. Note that both boundary conditions are in fact invariant under \(SU(N)\) gauge transformations \(F_{\mu\nu} \rightarrow U F_{\mu\nu} U^\dagger\), as they should be in order to make sense physically.  The plates are to be interpreted as physical 3D interfaces in a 4D bulk space, with the boundary conditions at the interfaces parameterizing their coupling to the bulk. We are thus \emph{not} considering boundary conditions coming from a variational principle for the GZ action. 

Let us write the left-hand side of Eq.~\eqref{eq:PEC-PMC} as $G_\mu^a$, such that we can handle both the PEC and PMC case simultaneously. We can lift the boundary conditions into the action by introducing Lagrange multiplier fields $(b_i^{-,a}, b_i^{+,a})$: 
\begin{align}\label{eq:SBC-1}
    S_\text{BC} = \int_{z=-\frac{L}2} d^3 \mathbf{x} \, b_i^{-,a} (\mathbf{x}) G_i^a + \int_{z=+\frac{L}2} d^3 \mathbf{x} \, b_i^{+,a} (\mathbf{x}) G_i^a.
\end{align}
Note that the auxiliary fields \(b^\pm\) only have three components \(i\in \{t,x,y\}\) (since $G_z^a$ vanishes automatically) and only depend on \(\mathbf{x} = (t,x,y)\). If we introduce the index notation \(\gamma\in \{-,+\}\) for the plates, we can take both terms together into an antisymmetric tensor structure \(H_{\mu\nu}^a\) (which differs for PEC and PMC):
\begin{align}\label{eq:H-PEC-PMC}
	H_{\mu \nu}^a = 
    \begin{cases}
	    \delta(z-z^\gamma) \, i \varepsilon_{\mu\nu i z} b_{i}^{\gamma, a} \quad &\text{(PEC)},\\
	    \delta(z-z^\gamma) \, (b_\mu^{\gamma, a} n_\nu - b_\nu^{\gamma, a} n_\mu)\quad &\text{(PMC)},
	\end{cases} 
\end{align}
in which we have defined \(b_z^\pm\equiv0\) and a summation over repeated plate indices is always understood unless explicitly stated otherwise.
With this tensor, the boundary action~\eqref{eq:SBC-1} can be written in a manifestly Lorentz covariant form as
\begin{align}\label{eq:SBC-2}
S_\text{BC} = \int \! d^4x \, \frac{1}{2} H_{\mu \nu}^a F_{\mu \nu}^a.
\end{align}
Since we are interested in investigating GZ theory in the presence of boundaries, from now on we consider the total action \(S\equiv S_\text{GZ}+S_\text{BC}\) and its associated functional integral
\begin{align}\label{eq:Z}
    Z = \int \! \mathcal{D}\tilde{\Phi} \, e^{-S},
\end{align}
where the auxiliary fields \(b^\pm\) are now also included in the functional measure. For the reader's convenience, we repeat the full action here:
\begin{align}\label{eq:action-full}
\begin{split}
    S = \int \!\! d^4x \, \bigg[ &\frac{1}{4} F_{\mu \nu}^a F_{\mu \nu}^a + \frac{1}{2} H_{\mu \nu}^a F_{\mu \nu}^a - \frac{\alpha}{2} \kappa^a \kappa^a+ \kappa^a \partial_\mu A_\mu^a + \bar{c}^a \partial_\mu D_\mu^{ab} c^b \\*
    &- \bar{\varphi}^{ac}_\mu \mathcal{M}^{ab}\varphi^{bc}_\mu + \bar{\omega}^{ac}_\mu \mathcal{M}^{ab} \omega^{bc}_\mu  + \gamma^2 g f^{abc} A_\mu^a (\varphi + \bar{\varphi})_\mu^{bc} -4\gamma^4(N^2-1) \bigg].
\end{split}
\end{align}


\subsection{Effective boundary action}\label{sec:3d-bnd-action}

An interesting approach to explore the Casimir effect within the functional integral methods of QFT is to integrate out all fields except the boundary multipliers \(b^\pm\), yielding a non-local three-dimensional boundary effective action that describes the dynamics on the plates~\cite{Dudal:2020yah, Canfora:2022xcx, Oosthuyse:2023mbs, Dudal:2024PEMC, Dudal:2024Robin, Dudal:2024DEM, Dudal2026}. We can start by integrating out the localizing fields \((\varphi,\bar \varphi)\) and \((\omega,\bar \omega)\), which effectively brings us back to the non-local formulation~\eqref{eq:action-GZ-non-local}. Since we are interested in the leading order contribution to the Casimir energy, we can safely restrict ourselves to the quadratic part of this action, ignoring interaction terms. Within this approximation, the inverse FP operator can be approximated by its leading order term, that is, $\left[\mathcal{M}^{-1}_{xy}\right]^{ad} \approx \frac{1}{(-\partial^2)} \delta^{ad} \delta(x-y)$. Thus, the quadratic part of the action of interest becomes
\begin{align}\label{eq:quad-action-GZ}
    S^{(2)} = \int \!\! d^4x \, \bigg[ &\frac{1}{4} \left(\partial_\mu A_\nu^a - \partial_\nu A_\mu^a\right)^2 + \frac{1}{2} H_{\mu \nu}^a \left(\partial_\mu A_\nu^a - \partial_\nu A_\mu^a\right) - \frac{\alpha}{2} \kappa^a \kappa^a+ \kappa^a \partial_\mu A_\mu^a + \bar{c}^a \partial^2 c^a + \gamma^4 g^2 N A_\mu^{a} \frac{1}{(-\partial^2)} A_\mu^{a} \bigg],
\end{align}
since we have $f^{abc} f^{aec} = N \delta^{be}$ in the horizon term. For now, we have dropped the vacuum term (the last term in~\eqref{eq:action-full}), since it will only become important when investigating the gap equation later on. The ghosts decouple in the quadratic approximation and the Nakanishi-Lautrup field $\kappa$ can be readily integrated, yielding
\begin{align}
    S^{(2)} =  \int \! d^4x \bigg[	\frac{1}{4} \left(\partial_\mu A_\nu^a - \partial_\nu A_\mu^a\right)^2  + H_{\mu \nu}^a \partial_\mu A_\nu^{a}  + \frac{1}{2 \alpha} (\partial_\mu A_\mu^a)^2  + \gamma^4 g^2 N A_\mu^{a} \frac{1}{(-\partial^2)} A_\mu^{a} \bigg].
\end{align}
Integrating by parts and grouping the terms, we obtain
\begin{align}
     S^{(2)} &= \int \! d^4x \left[	\frac{1}{2} A_\mu^a K_{\mu \nu}^{ab} A_\nu^b - (\partial_\mu H_{\mu \nu}^a) A_\nu^a \right],
\end{align}
where the quadratic operator $K_{\mu \nu}^{ab}$ in configuration space is given by
\begin{align}
    K_{\mu \nu}^{ab} = -\delta^{ab} \left[  \left(\partial^2 +  \frac{\lambda^4}{\partial^2} \right)\delta_{\mu \nu}- \left(1 - \frac{1}{\alpha} \right)\partial_\mu \partial_\nu  \right],
\end{align}
with the conventional shorthand
\begin{align}\label{lambdaGribov}
    \lambda^4 \equiv 2 \gamma^4 g^2 N
\end{align}
In Fourier space, this quadratic action translates to 
\begin{align}\label{eq:quad-action}
    S^{(2)} = \int \! \frac{d^4k}{(2\pi)^4}  \left[	\frac12 A_\mu^a(k) K_{\mu \nu}^{ab}(k) A_\nu^b(-k)  +v_\nu^a(k) A_\nu^a(-k) \right],
\end{align}
where we defined $v_\nu^a(k) \equiv i k_\mu H_{\mu \nu}^a(k)$, and the Fourier transform of the quadratic operator $K_{\mu \nu}^{ab}$ reads
\begin{align}
    K_{\mu \nu}^{ab}(k) = \delta^{ab} k^2 \left[  \left(\frac{k^4 + \lambda^4}{k^4} \right)\delta_{\mu \nu}- \left(1 - \frac{1}{\alpha} \right)\frac{k_\mu k_\nu}{k^2}  \right].
\end{align}
This operator has a well-defined inverse, that can be written as
\begin{align}\label{eq:K-inv}
    \left(K^{-1}\right)_{\mu \nu}^{ab} &= \frac{\delta^{ab}}{k^2} \left[ \left(\frac{k^4}{k^4 + \lambda^4}\right) \Theta_{\mu \nu} + \left(\frac{\alpha k^4}{k^4 + \alpha \lambda^4}\right) \Omega_{\mu \nu}  \right],
\end{align}
where $\Omega_{\mu\nu} \equiv \frac{k_\mu k_\nu}{k^2}$ and $\Theta_{\mu \nu} \equiv \delta_{\mu \nu} - \Omega_{\mu \nu}$ are the longitudinal and transverse projectors. In the Landau gauge $\left(\alpha \rightarrow 0\right)$, the gauge propagator simply becomes the transverse expression $\left(K^{-1}\right)_{\mu \nu}^{ab} = \delta^{ab} \left(\frac{k^2}{k^4 + \lambda^4}\right) \Theta_{\mu \nu}$. 

Performing the shift
\begin{align} \label{ShiftDecoupling}
 A_\mu^a(k) \rightarrow A_\mu^a(k) -  v_\rho^b(k) \left(K^{-1}\right)_{\mu \rho}^{ab}, 
\end{align}
which has a trivial Jacobian, one can see that the gauge field decouples from the boundary sector. Doing so, we find
\begin{align}
    S^{(2)} =  \int \! \frac{d^4k}{(2\pi)^4} &\bigg[ \frac{1}{2} A_\mu^a(k) K_{\mu \nu}^{ab} A_\nu^b(-k) - \frac{1}{2} v_\mu^a(k) \left(K^{-1}\right)_{\mu \nu}^{ab} v_\nu^b(-k) \bigg].
\end{align}
The functional integral thus factorizes into $Z = Z_A Z_b$, with the boundary functional integral given by
\begin{align}\label{eq:Zb}
    Z_b = \int \mathcal{D} b^+ \mathcal{D} b^- \, e^{-S_b},
\end{align}
in which the boundary effective action \(S_b\) is 
\begin{align}\label{eq:Sb}
    S_b = -\frac{1}{2} \! \int \!\!\frac{d^4k}{(2\pi)^4} \left[ k_\mu H_{\mu \rho}^a(k) \left(K^{-1}\right)_{\rho \sigma}^{ab} k_\nu H_{\nu \sigma}^b(-k) \right].
\end{align}
We can factorize out the boundary fields \(b^\pm\) by writing the antisymmetric tensor $H_{\mu \nu}$ in the following form:
\begin{align}
    H_{\mu \nu}^a(k) &= e^{i k_z z^\gamma} b_i^{\gamma, a}(\mathbf{k}) H_{i \mu \nu},
\end{align}
where $H_{i \mu \nu}$ is thus defined for each boundary condition as
\begin{align}\label{eq:HH-PEC-PMC}
	H_{i \mu \nu} = 
    \begin{cases}
	    i \varepsilon_{i \mu\nu z} \quad &\text{(PEC)}\\
	    \delta_{i \mu} n_\nu - \delta_{i \nu} n_\mu \quad &\text{(PMC)}.
	\end{cases}
\end{align}
This allows us to write the effective action in the form
\begin{align}\label{eq:Sb-GZ-form-KK}
    S_b = -\frac{1}{2} \int \frac{d^3\mathbf{k}}{(2\pi)^3} &\bigg\{  b_i^{\gamma, a}(\mathbf{k}) b_j^{\lambda, b}(\mathbf{-k}) \, \int \! \frac{dk_z}{2\pi} \left[  e^{i k_z (z^\gamma - z^\lambda)} k_\mu k_\nu H_{i \mu \rho} H_{j \nu \sigma}\left(K^{-1}\right)_{\rho \sigma}^{ab}  \right] \bigg\}.
\end{align}
Due to the antisymmetric nature of the tensor $H_{i \mu \nu}$, its contraction with the $\alpha$-dependent longitudinal part of $K^{-1}$ vanishes. Furthermore, the antisymmetry makes the term $\frac{k_\rho k_\sigma}{k^2}$ of the transverse projector vanish as well. After some algebra, one finds
\begin{align}
	k_\mu k_\nu H_{i \mu \rho} H_{j \nu \sigma} \left(K^{-1}\right)_{\rho \sigma}^{ab} = 
	    \delta^{ab}  \frac{k^2}{k^4+\lambda^4} \left[k^2 \delta_{ij}^\text{PMC} - \mathbf{k}^2 \Theta_{ij}(\mathbf{k})\right],
\end{align}
where the symbol \(\delta_{ij}^\text{PMC}\) equals \(\delta_{ij}\) in the PMC case, and is 0 in the PEC case. Here \(\Theta_{ij}(\mathbf{k})\) denotes the transversal projector in 3D momentum space, and from now on we understand projectors with Latin indices to live in 3D momentum space, i.e.,~\(\Theta_{ij}\equiv \Theta_{ij}(\mathbf{k})\), etc. To compute the \(k_z\)-integral in Eq.~\eqref{eq:Sb-GZ-form-KK}, we need the following decompositions:
\begin{align}
    \frac{k^2}{k^4 + \lambda^4} = \frac{1}{2} \left[\frac{1}{k^2 - i \lambda^2} + \frac{1}{k^2 + i \lambda^2}\right], \quad\text{and}\quad   \frac{k^4}{k^4 + \lambda^4} = \frac{i \lambda^2}{2} \left[\frac{1}{k^2 - i \lambda^2} - \frac{1}{k^2 + i \lambda^2}\right]+1.
\end{align}
Filling in the above expressions in Eq.~\eqref{eq:Sb-GZ-form-KK}, we find 
\begin{align}
    S_b = -\frac{1}{2} \int \!  \frac{d^3\mathbf{k}}{(2\pi)^3} \Bigg\lbrace b_i^{\gamma,a}(\mathbf{k})  \, b_j^{\lambda,a}(-\mathbf{k}) \int \! \frac{dk_z}{2\pi}  e^{i k_z(z^\gamma- z^\lambda)} \bigg[  \delta_{ij}^\text{PMC} \frac{i \lambda^2}{2} \left(\frac{1}{k^2 - i \lambda^2} - \frac{1}{k^2 + i \lambda^2}\right)
     - \frac12 \mathbf{k}^2 \Theta_{ij} \left(\frac{1}{k^2 + i \lambda^2} + \frac{1}{k^2 - i \lambda^2}\right)  \bigg]
 \Bigg\rbrace.
\end{align}
We have dropped the term $\int \! dk_z e^{i k_z (z^\gamma - z^\lambda)}\delta_{ij}^\text{PMC}$ since it is zero in PEC or if \(\gamma\neq \lambda\) or \(i\neq j\), and otherwise it equals $\delta(0)$, which also equals zero within dimensional regularization (see e.g.~\cite[Eq.~(4.2.6)]{Collins:1984xc} and \cite[below Eq.~(10.9)]{Zinn-Justin:2002ecy}). For a more detailed discussion about this, see~\cite{Dudal:2026uuv}.

Using the fact that for $L>0,\, a>0,\, b\in \mathbb{R}$ one has the integral identity
\begin{align}
    \int_{-\infty}^{\infty} \frac{dx}{2\pi} \, \frac{e^{-i x L}}{x^2 + a + i b }  = \frac{e^{-L \sqrt{a + i b}}}{2 \sqrt{a + i b}},
\end{align}
we can now compute the \(k_z\)-integral. This yields the effective boundary action 
\begin{align}\label{eq:Sb-quad}
    S_b &= -\frac{1}{2} \int \!  \frac{d^3\mathbf{k}}{(2\pi)^3} \Bigg\lbrace b_i^{\gamma,a}(\mathbf{k})  \,\mathbb{K}_{ij}^{\gamma\lambda}(\mathbf{k}) \,   b_j^{\lambda,a}(-\mathbf{k})  \Bigg\rbrace,
\end{align}
where the operator governing the dynamics of the boundary fields is given by
\begin{align}\label{eq:KK}
\begin{split}
   \mathbb{K}_{ij}^{\gamma\lambda}(\mathbf{k}) = &\frac{1}{4} \bigg[\left( i \lambda^2 \delta_{ij}^\text{PMC} - \mathbf{k}^2 \Theta_{ij} \right) \frac{e^{-|z^\gamma-z^\lambda|\sqrt{\mathbf{k}^2-i\lambda^2}}}{\sqrt{\mathbf{k}^2-i \lambda^2}} \\*
   &- \left( i \lambda^2 \delta_{ij}^\text{PMC} + \mathbf{k}^2 \Theta_{ij} \right) \frac{e^{-|z^\gamma-z^\lambda|\sqrt{\mathbf{k}^2+i\lambda^2}}}{\sqrt{\mathbf{k}^2+i \lambda^2}} \bigg].
\end{split}
\end{align}
Note that in the PEC case, the quadratic form is purely transversal, and thus not invertible. This corresponds to a residual boundary gauge freedom under \(b_i(\mathbf{k}) \rightarrow b_i(\mathbf{k}) + k_i f(\mathbf{k})\), for an arbitrary function \(f\), see~\cite{Dudal:2026uuv} for more details about this. Following the standard (perturbative) gauge-fixing procedure\footnote{Note that this is an approximation we are making here, as after all there could also be a  Gribov ambiguity for the dimensionally reduced 3D theory taking the full non-Abelian nature into account \cite{Dudal:2026uuv}. We will leave this for future work but already briefly comment about it in our outlook.} to deal with this issue as in \cite{Dudal:2020yah}, we will add the term \(-\frac{|\mathbf k|}2 \eta b_i^{\gamma,a} \Omega_{ij} b_j^{\gamma,a}\) to the boundary action to fix the residual gauge freedom for the PEC case:\footnote{The factor \(|\mathbf k|\) is included to make \(\eta\) a dimensionless gauge fixing parameter. As a consistency check, we will see later that all $\eta$-dependency will drop out.  } 
\begin{align}\label{eq:b-prop}
\begin{split}
   \mathbb{K}_{ij}^{\gamma\lambda}(\mathbf{k}) = &\frac{1}{4} \bigg[\left( i \lambda^2 \delta_{ij}^\text{PMC} - \mathbf{k}^2 \Theta_{ij} \right) \frac{e^{-|z^\gamma-z^\lambda|\sqrt{\mathbf{k}^2-i\lambda^2}}}{\sqrt{\mathbf{k}^2-i \lambda^2}}  \\*
   &- \left( i \lambda^2 \delta_{ij}^\text{PMC} + \mathbf{k}^2 \Theta_{ij} \right) \frac{e^{-|z^\gamma-z^\lambda|\sqrt{\mathbf{k}^2+i\lambda^2}}}{\sqrt{\mathbf{k}^2+i \lambda^2}} \bigg] + \eta |\mathbf k| \delta^\text{PEC}\Omega_{ij} \delta^{\gamma \lambda},
\end{split}
\end{align}
where it is understood that $\delta^{\rm PEC}$ is one for the PEC case and zero for the PMC case. That is, for the PMC case, \(\mathbb{K}\) is invertible and no such gauge-fixing term is required. We remark that $\mathbb{K}^{++}_{ij} = \mathbb{K}^{--}_{ij}$ and $\mathbb{K}^{+-}_{ij} = \mathbb{K}^{-+}_{ij}$.

For future use, it will be helpful to explicitly calculate the inverse of \(\mathbb{K}^{\gamma\lambda}_{ij} = \begin{pmatrix}
    \mathbb{K}^{++}_{ij} & \mathbb{K}^{-+}_{ij} \\
    \mathbb{K}^{-+}_{ij} & \mathbb{K}^{++}_{ij}
\end{pmatrix}\). Let us succinctly write 
\begin{align}
\begin{split}
    \mathbb{K}^{++}_{ij} = x \Theta_{ij} + y \Omega_{ij},\\
    \mathbb{K}^{-+}_{ij} = u \Theta_{ij} + v \Omega_{ij}.
\end{split}
\end{align}
These coefficients can be read off from Eq.~\eqref{eq:b-prop}, but for the reader's convenience, we state them explicitly here: 
\begin{align}
    y &= \delta^\text{PMC}\frac{i\lambda^2}4 \left( \frac{1}{\sqrt{\mathbf{k}^2-i\lambda^2}} -\frac{1}{\sqrt{\mathbf{k}^2+i\lambda^2}}\right) + \eta|\mathbf k|\delta^\text{PEC},\\*
    x &= \frac{-\mathbf{k}^2}4 \left( \frac{1}{\sqrt{\mathbf{k}^2-i\lambda^2}} +\frac{1}{\sqrt{\mathbf{k}^2+i\lambda^2}}\right)+y\delta^\text{PMC},\\*
    v &= \delta^\text{PMC}\frac{i\lambda^2}4 \left(\frac{ e^{-L \sqrt{\mathbf{k}^2-i \lambda ^2}}}{\sqrt{\mathbf{k}^2-i \lambda ^2}} -\frac{ e^{-L \sqrt{\mathbf{k}^2+i \lambda ^2}}}{\sqrt{\mathbf{k}^2+i \lambda ^2}}\right),\\*
    u &= \frac{-\mathbf{k}^2}{4} \left(\frac{e^{-L \sqrt{\mathbf{k}^2-i \lambda ^2}}}{\sqrt{\mathbf{k}^2-i \lambda ^2}}+\frac{e^{-L \sqrt{\mathbf{k}^2+i \lambda ^2}}}{\sqrt{\mathbf{k}^2+i \lambda ^2}}\right)+v.
\end{align}
One can check that the operator $\mathbb{K}^{-1}$ can then be written as
\begin{align}\label{bfprop}
    (\mathbb{K}^{-1})_{ij}^{\gamma\lambda} = E^{\gamma\lambda}\Theta_{ij} +F^{\gamma\lambda} \Omega_{ij}, 
\end{align}
in which
\begin{align}
    E^{\gamma\lambda} = \frac{1}{x^2-u^2}\begin{pmatrix}
        x & -u \\
        -u & x
    \end{pmatrix}, \quad \text{and} \quad
    F^{\gamma\lambda} = \frac{1}{y^2-v^2}\begin{pmatrix}
        y & -v \\
        -v & y
    \end{pmatrix}.
\end{align}


\subsection{The generating functional $Z[J]$ and the boundary-modified gluon propagator}\label{sec:A-prop}

The presence of boundaries in spacetime, such as the parallel plates considered here, significantly influences the behavior of quantum fields, leading to modifications of fundamental quantities like the gluon and ghost propagators. To analyze the impact of the non-trivial boundary conditions considered here on the gluon propagator, let us consider the generating functional in the presence of external sources:
\begin{align}
    Z[J] = \int \! \mathcal{D}\!A \, \mathcal{D} b^+ \mathcal{D} b^- \, e^{-S[J]},
\end{align}
where we have added a source term to the quadratic action \eqref{eq:quad-action} 
\begin{align}
    S[J] = \int \! \frac{d^4k}{(2 \pi)^4}  \left[	\frac{1}{2} A_\mu^a(k) K_{\mu \nu}^{ab} A_\nu^b(-k)  +v_\nu^a(k) A_\nu^a(-k) \right] -  \int \! \frac{d^4k}{(2 \pi)^4} J_\mu^a(k) A_\mu^a(-k).
\end{align}
One can rewrite the above action in a more convenient form:
\begin{align}
    S[J] = \int \! \frac{d^4k}{(2 \pi)^4} \left[	 \frac{1}{2} A_\mu^a(k) K_{\mu \nu}^{ab} A_\nu^b(-k)  -\mathcal{J}_\nu^a(k)A_\nu^a(-k) \right], 
\end{align}
where we defined 
\begin{align}
    \mathcal{J}_\nu^a(k) \equiv J_\nu^a(k) - v_\nu^a(k). 
\end{align}
Now, we can propose the following shift:
\begin{align}
A_\mu^a(k) \rightarrow A_\mu^a(k) + \mathcal{J}_\rho^b(k) \left(K^{-1}\right)_{\mu \rho}^{ab}.
\end{align}
Thus, we will get
\begin{align}
    S[J] = \int \! \frac{d^4k}{(2 \pi)^4} \left[	 \frac{1}{2} A_\mu^a(k) K_{\mu \nu}^{ab} A_\nu^b(-k)  - \frac{1}{2} \mathcal{J}_\mu^a(k) \left(K^{-1}\right)_{\mu \nu}^{ab} \mathcal{J}_\nu^b(-k) \right]. 
\end{align}
As such, we have decoupled the gluon and boundary fields, and we can rewrite the functional integral as
\begin{align}
\begin{split}
    Z[J] &= \int \! \mathcal{D}A e^{-\int \! \frac{d^4k}{(2 \pi)^4} \frac{1}{2} A_\mu^a(k) K_{\mu \nu}^{ab} A_\nu^b(-k)} \int \! \mathcal{D}b^+ \mathcal{D}b^- \, e^{\int \! \frac{d^4k}{(2 \pi)^4} \frac{1}{2} \mathcal{J}_\mu^a(k) \left(K^{-1}\right)_{\mu \nu}^{ab} \mathcal{J}_\nu^b(-k)}  \\*
    &= \left[\det\left(K\right)\right]^{-1/2} \int \! \mathcal{D}b^+ \mathcal{D}b^- \, e^{\int \! \frac{d^4k}{(2 \pi)^4} \frac{1}{2} \mathcal{J}_\mu^a(k) \left(K^{-1}\right)_{\mu \nu}^{ab} \mathcal{J}_\nu^b(-k)}.
\end{split}
\end{align}
The integrand inside the second exponential can be rewritten as
\begin{align}\label{eq:quad-sources}
\begin{split}
    \frac{1}{2} \mathcal{J}_\mu^a(k) \left(K^{-1}\right)_{\mu \nu}^{ab} \mathcal{J}_\nu^b(-k) &= \frac{1}{2} \bigg[ J_\mu^a(k) \left(K^{-1}\right)_{\mu \nu}^{ab} J_\nu^b(-k) +  v_\mu^a(k) \left( K^{-1} \right)_{\mu \nu}^{ab} v_\nu^b(-k) \\*
    &- v_\mu^a(k) \left( K^{-1} \right)_{\mu \nu}^{ab} J_\nu^b(-k) - J_\mu^a(k) \left( K^{-1} \right)_{\mu \nu}^{ab} v_\nu^b(-k)\bigg].
\end{split}
\end{align}
The first term only involves the sources. For the second term, we already know that, if we define, 
\begin{align}
    S_b = -\frac{1}{2} \int \! \frac{d^4k}{(2 \pi)^4} \left[ v_\mu^a(k) \left( K^{-1} \right)_{\mu \nu}^{ab} v_\nu^b(-k)\right],
\end{align}
after some computational effort we can obtain \(\mathbb{K}\), cf.~\eqref{eq:KK}, and write
\begin{align}
    S_b &= -\frac{1}{2} \int \!  \frac{d^3\mathbf{k}}{(2\pi)^3} \Bigg\lbrace b_i^{\gamma,a}(\mathbf{k})  \,\mathbb{K}_{ij}^{\gamma\lambda}(\mathbf{k}) \,  b_j^{\lambda,1}(-\mathbf{k})  \Bigg\rbrace.
\end{align}
For the third term of \eqref{eq:quad-sources}, we can use $v_\mu^a(k) = i k_\rho H_{\rho \mu}^a(k)$, with $H_{\rho \mu}^a(k) = e^{i k_z z^\gamma} b_i^{\gamma,a}(\mathbf{k}) H_{i\rho \mu}$, cf.~\eqref{eq:HH-PEC-PMC}, to obtain:
\begin{align}
    \int \! \frac{d^4k}{(2 \pi)^4} \left[-v_\mu^a(k) \left( K^{-1} \right)_{\mu \nu}^{ab} J_\nu^b(-k)\right] &= \int \! \frac{d^4k}{(2 \pi)^4}  \left[-i e^{i k_z z^\gamma} \frac{k^2}{k^4+\lambda^4}k_\mu b_i^{\gamma,a}(\mathbf{k})H_{i\mu\nu}J_\nu^a(-k)\right] = \int \! \frac{d^3\mathbf{k}}{(2 \pi)^3}  b_i^{\gamma,a}(\mathbf{k}) W_i^{\gamma, a}(-\mathbf{k}),
\end{align}
where we naturally defined 
\begin{align}\label{eq:def-W}
   W_i^{\gamma, a}(-\mathbf{k}) \equiv \int \! \frac{dk_z}{2 \pi}\left[ -i e^{i k_z z^\gamma}  \frac{k^2}{k^4 + \lambda^4} k_\mu H_{i\mu\nu} J_\nu^a(-k)  \right].
\end{align}
Note that once more, the gauge parameter \(\alpha\) has dropped out. Substituting \(\mathbf k\rightarrow -\mathbf k\), one can see that the last term of Eq.~\eqref{eq:quad-sources} equals the third. 
Gathering all four terms, we find
\begin{align}
     Z[J] = \left[\det\left(K\right)\right]^{-1/2} e^{\int \! \frac{d^4k}{(2 \pi)^4} \left[ \frac{1}{2} J_\mu^a(k) \left(K^{-1}\right)_{\mu \nu}^{ab} J_\nu^b(-k)\right] }  \int \! \mathcal{D}b^+ \mathcal{D}b^- \, e^{\int \!  \frac{d^3\mathbf{k}}{(2\pi)^3} \left[ \frac12 b_i^{\gamma,a}(\mathbf{k})  \,\mathbb{K}_{ij}^{\gamma\lambda} \,  b_j^{\lambda,a}(-\mathbf{k}) +  b_i^{\gamma,a}(\mathbf{k}) W_i^{\gamma, a}(-\mathbf{k}) \right]}.
\end{align}
We can now decouple the \(b\)-field from the source using the following shift
\begin{align}
    b_i^{a, \gamma} \rightarrow b_i^{a, \gamma} - \left(\mathbb{K}^{\gamma\lambda}\right)^{-1}_{ij} W_j^{\lambda, a}.
\end{align}
Thus, we get
\begin{align}
    \frac12 b_i^{\gamma,a}  \,\mathbb{K}_{ij}^{\gamma\lambda} \,  b_j^{\lambda,a} +  b_i^{\gamma,a} W_i^{\gamma, a} \longrightarrow  \frac12 b_i^{\gamma,a}  \,\mathbb{K}_{ij}^{\gamma\lambda} \,  b_j^{\lambda,a} - \frac{1}{2} W_i^{\gamma, a} \left(\mathbb{K}_{ij}^{\gamma\lambda}\right)^{-1}  W_j^{\lambda, a}.
\end{align}
Therefore, the boundary-modified generating functional in the presence of external sources becomes
\begin{align}
     Z[J] &= \left[\det K \right]^{-1/2} e^{\int \! \frac{d^4k}{(2 \pi)^4} \left[ \frac{1}{2} J_\mu^a(k) \left(K^{-1}\right)_{\mu \nu}^{ab} J_\nu^b(-k) \right]} e^{\int \!  \frac{d^3\mathbf{k}}{(2\pi)^3} \left[ -\frac{1}{2} W_i^{\gamma, a}(\mathbf{k}) \left(\mathbb{K}_{ij}^{\gamma\lambda}\right)^{-1}  W_j^{\lambda, a}(-\mathbf{k}) \right]}  \int \! \mathcal{D}b^+ \mathcal{D}b^- \, e^{\int \!  \frac{d^3\mathbf{k}}{(2\pi)^3} \left[ \frac12 b_i^{\gamma,a}(\mathbf{k})  \,\mathbb{K}_{ij}^{\gamma\lambda} \,  b_j^{\lambda,a}(-\mathbf{k})\right]} \nonumber \\*
     &= \left[\det K \right]^{-1/2}  \left[\det \mathbb{K} \right]^{-1/2} e^{\int \! \frac{d^4k}{(2 \pi)^4} \left[ \frac{1}{2} J_\mu^a(k) \left(K^{-1}\right)_{\mu \nu}^{ab} J_\nu^b(-k) \right]} e^{\int \!  \frac{d^3\mathbf{k}}{(2\pi)^3} \left[ -\frac{1}{2} W_i^{\gamma, a}(\mathbf{k}) \left(\mathbb{K}_{ij}^{\gamma\lambda}\right)^{-1}  W_j^{\lambda, a}(-\mathbf{k}) \right]}.
\end{align}
The boundary-modified gluon propagator can be obtained by taking the double derivative of \(\ln Z\) with respect to the sources:
\begin{align}
    \langle  A_\mu^a(p)A_\nu^b(q)\rangle &= \frac{\delta^2}{\delta J_\mu^a(p)\delta J_\nu^b(q)} \ln Z[J]{\Big|_{J=0}} \\ 
    &= \frac{\delta^2}{\delta J_\mu^a(p)\delta J_\nu^b(q)}  \left\{ \int \! \frac{d^4k}{(2 \pi)^4} \left[ \frac{1}{2} J_\mu^a(k) \left(K^{-1}\right)_{\mu \nu}^{ab} J_\nu^b(-k) \right] + \int \!  \frac{d^3\mathbf{k}}{(2\pi)^3} \left[ -\frac{1}{2} W_i^{\gamma, c}(\mathbf{k}) \left(\mathbb{K}_{ij}^{\gamma\lambda}\right)^{-1}  W_j^{\lambda, c}(-\mathbf{k}) \right] \right\}.
    \end{align}
The first term is already present in the case without plates, but we find an extra term related to the existence of a boundary. Thus, the boundary-modified gluon propagator can be written as:
\begin{align}
     \langle  A_\mu^a(p)A_\nu^b(q)\rangle = \delta^{(4)}(p+q) \left(K^{-1}\right)_{\mu \nu}^{ab} + D_{\mu\nu}^{ab}(p,q),\label{eq:A-prop}
\end{align}
where \(K^{-1}\) is given by \eqref{eq:K-inv}, and the boundary correction by
\begin{align}
    D_{\mu\nu}^{ab}(p,q) =-\frac{1}{2} \int \!  \frac{d^3\mathbf{k}}{(2\pi)^3} \left[ \frac{ \delta W_i^{\gamma, c}(\mathbf{k})}{\delta J_\mu^a(p)} \left(\mathbb{K}^{-1}\right)_{ij}^{\gamma\lambda}  \frac{\delta W_j^{\lambda, c}(-\mathbf{k})}{\delta J_\nu^b(q)} \right].
\end{align}
Keeping in mind the definition of \(W\) in~\eqref{eq:def-W}, we have
\begin{equation}
    \frac{ \delta W_i^{\gamma, c}(\mathbf{k})}{\delta J_\mu^a(p)} = \int \! \frac{dk_z}{2 \pi}\left[ i e^{-i k_z z^\gamma}  \left(\frac{k^2}{k^4 + \lambda^4}\right)   k_\alpha H_{i\alpha\mu} \delta^{ac} \delta^{(4)}(p-k)  \right],
\end{equation}
the modification in the gluon propagator caused by the presence of boundaries reads
\begin{align}\label{GluonPropModif}
    D_{\mu\nu}^{ab}(p,q) = \frac12 \delta^{ab} \delta^{(3)}(\mathbf p + \mathbf q)e^{-i(p_z z^\gamma +q_z z^\lambda)} \frac{p^2}{p^4+\lambda^4} \frac{q^2}{q^4+\lambda^4}  p_\alpha H_{i\alpha\mu} \left(\mathbb{K}^{-1}\right)_{ij}^{\gamma\lambda} q_\beta H_{j\beta\nu}.
\end{align}
It is worth pointing out that the boundary modification presented above comes with a three-dimensional delta function instead of the usual four-dimensional one. This is due to the broken translation invariance in the $z$-direction, perpendicular to the plates. 


\section{Computing the Casimir energy}\label{sec:free-Gribov-param}

The Casimir energy of the system considered here can be defined as the energy difference between the system in the presence of plates (boundaries) and without them (free space). A practical way of obtaining the latter from the former is simply to take the limit $L \rightarrow \infty$. Note that any $L$-independent contribution to the vacuum energy would vanish when taking this difference. Therefore, we can simply discard $L$-independent contributions when computing the Casimir energy because such terms will not contribute to it.

In Sec.~\ref{sec:3d-bnd-action} we have factorized the functional integral~\eqref{eq:Z} into two independent contributions under the free field approximation adopted here: a purely gluonic part $Z_A$ and a boundary contribution $Z_b$, cf.~\eqref{eq:Zb}. Thus, in principle, the Casimir energy receives two distinct contributions: one arising from the gluons in the bulk, and another from the auxiliary fields on the boundary. In the context of pure YM theory, there can be no $L$-dependence in the gluon part $Z_A$ after the shift~\eqref{ShiftDecoupling}, guaranteeing that its contribution to the Casimir effect can be safely neglected. However, within the GZ framework, the situation is more subtle. The Gribov parameter $\lambda$, which enters $Z_A$, is dynamically determined through the gap equation~\eqref{VacuumGapEquation}. Since this equation features the vacuum energy, which clearly depends on the plate separation $L$, it is plausible that $\lambda$ might acquire a dependence on $L$ as well. This implies that $Z_A$ could exhibit an indirect $L$-dependence through $\lambda(L)$, as determined by the gap equation. Nevertheless, it seems reasonable to expect that such a dependence would constitute only a sub-leading effect, justifying the omission of a possible $L$-dependence of $Z_A$ as a first approximation. However, as we will discuss later, the situation is even simpler: we can show that the presence of plates does not actually modify the gap equation within the current approximation, thus not introducing any $L$-dependence in the Gribov parameter. Therefore, it is justified to consider $\lambda$ as $L$-independent, allowing us to safely ignore contributions to the Casimir energy coming from $Z_A$.

In the following, we will calculate the Casimir energy using two different approaches. The first approach computes it directly from the functional integral, while the second one starts from the energy-momentum tensor. 

\subsection{Casimir energy via the functional integral}\label{sec:Casimir-direct}

We can compute the Casimir energy through the functional integral by using the identity \cite[Sec.~III]{Dudal:2024DEM}
\begin{equation}
    Z = Z_A Z_b = \int \mathcal{D}A \,e^{-\frac12\int\frac{d^4k}{(2\pi)^4} A_\mu^a(k) K_{\mu\nu}^{ab} A_\nu^b(-k)}  \int \! \mathcal{D}b^+ \mathcal{D}b^- \, e^{-S_b}  = e^{-\ell_t E_\text{bulk}} e^{-\ell_t E_\text{bnd}},
\end{equation}
with \(\ell_t\) the (infinite) length of the integration volume in the \(t\)-direction, and \(E_\text{bulk}\) (resp.~\(E_\text{bnd}\)) the vacuum energy of the bulk (resp.~boundary) theory.
Since we will show that \(Z_A\) is independent of \(L\), only the boundary theory contributes to the Casimir energy density, which is given by 
\begin{equation}
    \mathcal{E}_\text{Cas} = E_\text{bnd}/\ell_x \ell_y = - \frac{\ln Z_b}{\ell_t\ell_x\ell_y}.    
\end{equation}
As we have already brought $S_b$ in the quadratic form~\eqref{eq:Sb-quad}, we can directly compute $Z_b$ with a Gaussian integration
\begin{align}
    Z_b = \int \! \mathcal{D}b^+ \mathcal{D}b^- \, e^{-S_b} = C \left( \text{Det}\, \mathbb{K} \right)^{-1/2},
\end{align}
where $C$ is an irrelevant infinite factor that we will not continue writing. The determinant \(\text{Det}\, \mathbb{K}\) must be understood in the functional sense (i.e.~both in the continuous index $\mathbf{k}$ and in the discrete indices), meaning the product of all the eigenvalues of $\mathbb{K}$. We have
\begin{align}
    \mathcal{E}_\text{Cas} = \frac{1}{2} \frac{1}{\ell_t\ell_x\ell_y}\log\left(\text{Det}\, \mathbb{K}\right).
\end{align}
We now use the well-known identity $\log \det A= \text{tr} \log A$ with respect to the continuous index, giving us
\begin{align}
    \mathcal{E}_\text{Cas} = \frac12  \frac{1}{\ell_t\ell_x\ell_y} \text{Tr}_\mathbf{k} \log \big( \det \mathbb{K}(\mathbf k) \big) =\frac{1}{2} \int \frac{d^3 \mathbf{k}}{(2\pi)^3} \log \big( \det \mathbb{K}(\mathbf k) \big),
\end{align}
where $\det \mathbb{K}(\mathbf k)$ is only the discrete matrix determinant, and the volume factor \(1/\ell_t\ell_x\ell_y\) has canceled with the volume coming from the \(\mathbf{k}\)-trace. We thus need to compute the matrix determinant $\det \mathbb{K}(\mathbf k)$. Let us repeat the boundary dynamical operator \(\mathbb{K}(\mathbf{k})\) here for the reader's convenience:
\begin{align}
\begin{split}
   \mathbb{K}_{ij}^{\gamma\lambda}(\mathbf{k}) = &\frac{1}{4} \bigg[\left( i \lambda^2 \delta_{ij}^\text{PMC} - \mathbf{k}^2 \Theta_{ij} \right) \frac{e^{-|z^\gamma-z^\lambda|\sqrt{\mathbf{k}^2-i\lambda^2}}}{\sqrt{\mathbf{k}^2-i \lambda^2}}  \\*
   &- \left( i \lambda^2 \delta_{ij}^\text{PMC} + \mathbf{k}^2 \Theta_{ij} \right) \frac{e^{-|z^\gamma-z^\lambda|\sqrt{\mathbf{k}^2+i\lambda^2}}}{\sqrt{\mathbf{k}^2+i \lambda^2}} \bigg] + \eta |\mathbf k|\delta^\text{PEC}\Omega_{ij} \delta^{\gamma \lambda}.
\end{split}
\end{align}
We can picture this as a \(2\times2\) block matrix consisting of \(3\times3\) blocks: \(\mathbb{K}^{\gamma\lambda}_{ij} = \begin{pmatrix}
    \mathbb{K}^{++}_{ij} & \mathbb{K}^{-+}_{ij} \\
    \mathbb{K}^{-+}_{ij} & \mathbb{K}^{++}_{ij}
\end{pmatrix}\). For such a block matrix, when $\mathbb{K}^{++}$ is invertible, one has the general identity \cite{powell2011}
\begin{equation}
    \det \mathbb{K}(\mathbf{k}) = \det \mathbb{K}^{++} \cdot \det \left( \mathbb{K}^{++} - \mathbb{K}^{-+} \left(\mathbb{K}^{++}\right)^{-1} \mathbb{K}^{-+} \right).
\end{equation}
Since both \(\mathbb{K}^{++}\) and \(\mathbb{K}^{-+}\) are linear combinations of the longitudinal and transversal projector, these matrix products and determinants can easily be computed. We shall do so separately for the PMC and PEC case.

\subsubsection{PMC case}

Computing the determinant above for the PMC case, one finds
\begin{align}
    \det \mathbb{K}^\text{PMC}(\mathbf{k}) = \frac{-\lambda^4}{4096 \,\bar \omega^2 \omega^2} \times\left[(\bar \omega - \omega)^2 - \left(\bar \omega e^{-L \omega} - \omega e^{-L \bar \omega} \right)^2\right] 
    \times \left[(\bar \omega + \omega)^2 - \left(\bar \omega e^{-L \bar \omega} + \omega e^{-L \omega}\right)^2\right]^2,
\end{align}
where we have adopted the shorthand notation
\begin{align}
    \begin{split}
        \omega \equiv \sqrt{\mathbf{k}^2 + i \lambda^2}, \\*
        \bar \omega \equiv \sqrt{ \mathbf{k}^2 - i \lambda^2}.
    \end{split}
\end{align}

To compute the Casimir energy, we have to take the logarithm of the above determinant. The terms that do not depend directly on $L$ are irrelevant for our purposes\footnote{Here we are only considering the direct \(L\)-dependence. A possible indirect dependence through the Gribov mass will be considered later.} and will be discarded. Thus, we get
\begin{align}
\log \det \mathbb{K}^\text{PMC} = \log\left[1 - \frac{(\bar \omega e^{-L \omega} - \omega e^{-L \bar \omega} )^2}{(\bar \omega - \omega)^2}\right]+2 \log \left[1 - \frac{(\bar \omega e^{-L \bar \omega} + \omega e^{-L \omega} )^2}{(\bar \omega + \omega)^2} \right].
\end{align}
Therefore, the Casimir energy is given by
\begin{equation}\label{eq:Cas-energy-PMC}
    \mathcal{E}_\text{Cas}^\text{PMC} = \mathcal{E}_1^\text{PMC} + 2\mathcal{E}_2^\text{PMC},
\end{equation}
with
\begin{align}
     \mathcal{E}_1^\text{PMC} &= \frac{(N^2-1)}{4 \pi^2} \int_0^\infty dk\, k^2 \log \left[1 - \frac{(\bar \omega e^{-L \omega} - \omega e^{-L \bar \omega} )^2}{(\bar \omega - \omega)^2}\right],  \\*
     \mathcal{E}_2^\text{PMC} &= \frac{(N^2-1)}{4 \pi^2} \int_0^\infty dk \, k^2 \log \left[1 - \frac{(\bar \omega e^{-L \bar \omega} + \omega e^{-L \omega} )^2}{(\bar \omega + \omega)^2} \right].
\end{align}
These expressions for \(\mathcal{E}_1^\text{PMC}\) and \(\mathcal{E}_2^\text{PMC}\) can also be cast in a manifestly real form by introducing the real variables $x=\frac{\omega + \bar \omega}{2}$ and $y = \frac{\omega - \bar \omega}{2i}$.
In these variables, the energy expressions become
\begin{align}
     \mathcal{E}_1^\text{PMC} &= \frac{(N^2-1)}{4 \pi^2}  \int_0^\infty dk\, k^2 \log \bigg[1 - \bigg( e^{-L x} \Big(  \cos(L y) + \frac{x}{y} \sin(L y) \Big) \bigg)^2\bigg],  \\*
     \mathcal{E}_2^\text{PMC} &= \frac{(N^2-1)}{4 \pi^2}  \int_0^\infty dk \, k^2 \log \bigg[1 - \bigg( e^{-L x} \Big( \cos (L y) + \frac{y}{x} \sin (L y)\Big)  \bigg)^2\bigg].
\end{align}

\begin{figure}
    \centering
    \subfloat[\centering \(\mathcal{E}_\text{Cas}^\text{PMC}(L)\)]{{\includegraphics[width=0.45\linewidth]{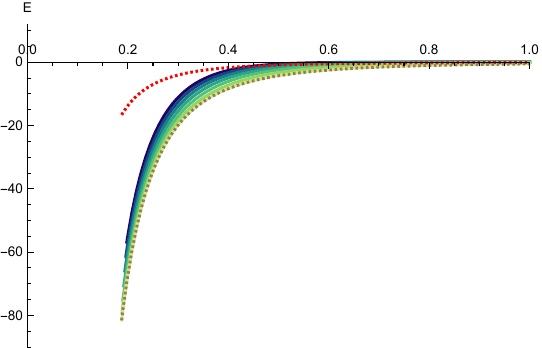}}}
    \qquad
    \subfloat[\centering Log-log plot of \(-\mathcal{E}_\text{Cas}^\text{PMC}(L)\)]{{\includegraphics[width=0.45\linewidth]{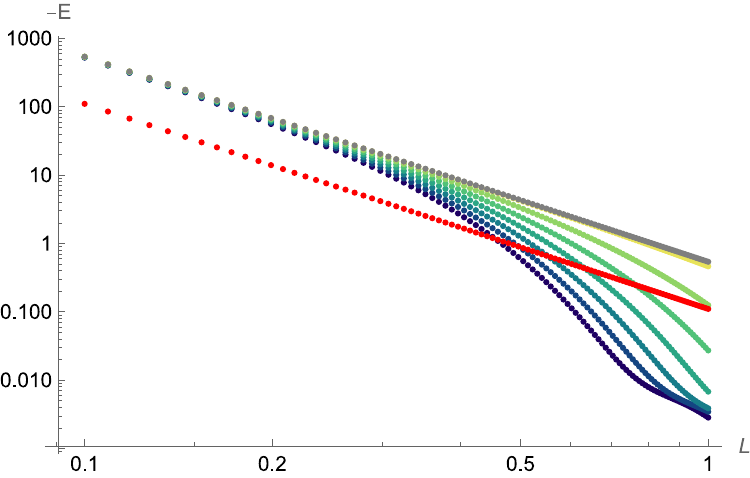}}}
    \caption{The Casimir energy density \eqref{eq:Cas-energy-PMC} for PMC plates as a function of \(L\in[0,1]\). Different values of \(\lambda^2 \in \{ 1,6,11,\dots, 31\}\) are represented by a gradient that is darker for larger values. The limit \(\lambda\rightarrow0\) is shown in gray dots, and the Yang-Mills Casimir energy \eqref{eq:energy-YM} is shown in red dots for reference. All numerical values are expressed in units \(\LambdaMS = 1\). }
    \label{fig:energy-PMC}
\end{figure}

We have not been able to find an analytic solution for these integrals, but they can be solved numerically for given values of \(L\) and \(\lambda\). 
In Fig.~\ref{fig:energy-PMC}, the Casimir energy~\eqref{eq:Cas-energy-PMC} for PMC boundaries is shown as a function of \(L\) for several distinct values of \(\lambda\), together with the known YM result for comparison. 

We see that the PMC Casimir energy for GZ is negative, but substantially bigger (in absolute terms) than for Yang-Mills, especially for small values of \(\lambda\). 
The corresponding Casimir force is attractive for all \(\lambda\)'s.
From the log-log plot, we can read off that for small \(\lambda\), the GZ Casimir energy density approaches a power law \(1/L^3\), but with a larger coefficient (in absolute terms) than in Yang-Mills (cfr.~Sec.~\ref{sec:lambda0}). The bigger the value of \(\lambda\), the more the Casimir energy exhibits an exponential behavior instead of a power law. The limit $\lambda \to 0$ is more subtle and requires a separate discussion. This will be done in Sec.~\ref{sec:lambda0}.

\subsubsection{PEC case}

Considering now the PEC case, one finds for the determinant the following result:
\begin{align}
    \det \mathbb{K}^\text{PEC}(\mathbf{k}) &= \frac{\eta^2\mathbf k^{10}}{256 \bar \omega^4 \omega^4} 
    \left[(\bar \omega + \omega)^2 - (\bar \omega e^{-L \omega} + \omega e^{-L \bar \omega} )^2\right]^2.
\end{align}
Taking the logarithm and discarding the parts independent of $L$, we get
\begin{align}
\log \det \mathbb{K}^\text{PEC} &= 2 \log \left[1 - \frac{(\bar \omega e^{-L \omega} + \omega e^{-L \bar \omega} )^2}{(\bar \omega + \omega)^2} \right].
\end{align}
Therefore, the Casimir energy is given by
\begin{equation}\label{eq:Cas-energy-PEC}
    \mathcal{E}_\text{Cas}^\text{PEC} = 2\mathcal{E}^\text{PEC}_2 = 2\frac{(N^2-1)}{4 \pi^2} \int_0^\infty dk\, k^2 \log \left[1 - \frac{(\bar \omega e^{-L \omega} + \omega e^{-L \bar \omega} )^2}{(\bar \omega + \omega)^2} \right].
\end{equation}
This expression can also be cast in a manifestly real form 
\begin{align}\label{eq:Cas-energy-PEC-real}
     \mathcal{E}^\text{PEC}_2 &= \frac{(N^2-1)}{4 \pi^2} \int_0^\infty dk\, k^2 \log \bigg[1 - \bigg( e^{-L x} \Big(  \cos(L y) - \frac{y}{x} \sin(L y) \Big) \bigg)^2\bigg].
\end{align}

Notice the relative sign difference between \(\mathcal{E}_2^\text{PMC}\) and \(\mathcal{E}_2^\text{PEC}\). Furthermore, note that the residual gauge-fixing procedure in the PEC case has ``eaten away'' the energy contribution \(\mathcal{E}_1\). To acquire a better understanding of this phenomenon, it is useful to consider what happens in the limit \(\lambda\rightarrow0\), which we will do in the next subsection.

\begin{figure}
    \centering
    \subfloat[\centering \(\mathcal{E}_\text{Cas}^\text{PEC}(L)\)]{{\includegraphics[width=0.45\linewidth]{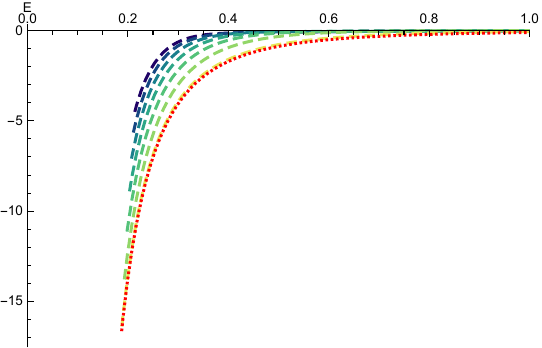}}}
    \qquad
    \subfloat[\centering Log-log plot of \(-\mathcal{E}_\text{Cas}^\text{PEC}(L)\)]{{\includegraphics[width=0.45\linewidth]{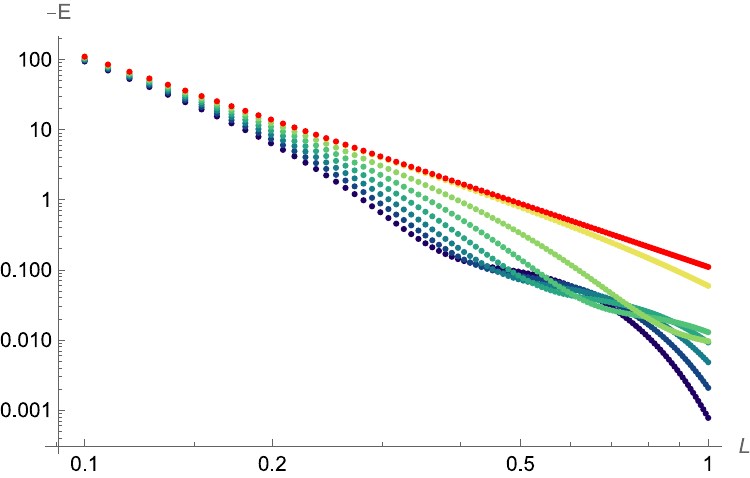}}}
    \caption{The Casimir energy density \eqref{eq:Cas-energy-PEC} for PEC plates as a function of \(L\in[0,1]\). Different values of \(\lambda^2 \in \{ 1,6,11,\dots, 31\}\) are represented by a gradient that is darker for larger values. The limit \(\lambda\rightarrow0\) coincides with the Yang-Mills Casimir energy \eqref{eq:energy-YM}, and is shown in red dots for reference. All numerical values are expressed in units \(\LambdaMS = 1\). }
    \label{fig:energy-PEC}
\end{figure}

We have not been able to find an analytic solution for this integral either, but it can be solved numerically for given values of \(L\) and \(\lambda\). In Fig.~\ref{fig:energy-PEC}, the Casimir energy \eqref{eq:Cas-energy-PEC} for PEC boundaries is shown as a function of \(L\) for several distinct values of \(\lambda\), together with the known YM result for comparison.

We see that, for large values of \(\lambda\), the PEC Casimir energy for GZ is substantially smaller (in absolute terms) than for YM.
However, for smaller values of \(\lambda\), the PEC Casimir energy for GZ approaches the YM Casimir energy.
Here as well, the corresponding Casimir force is attractive for all \(\lambda\)'s.
From the log-log plot as well, we can read off that for small \(\lambda\), the GZ Casimir energy density approaches the same \(1/L^3\) power law as in YM (cfr.~Sec.~\ref{sec:lambda0}). 
The bigger the value of \(\lambda\), the more the Casimir energy exhibits an exponential behavior instead of a power law. 

\begin{figure}
    \centering
    \includegraphics[width=0.6\linewidth]{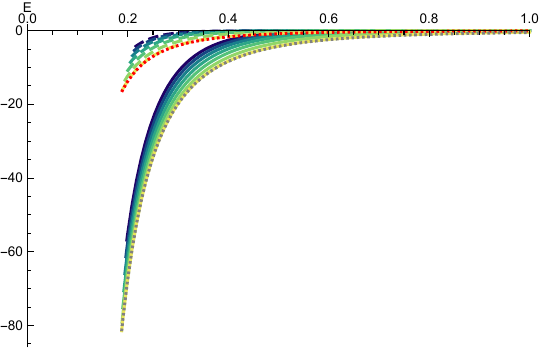}
    \caption{The Casimir energy density \eqref{eq:Cas-energy-PMC} for PMC plates and the Casimir energy \eqref{eq:Cas-energy-PEC} for PEC plates (dashed) as a function of \(L\in[0,1]\). Different values of \(\lambda^2 \in \{ 1,6,11,\dots, 31\}\) are represented by a gradient that is darker for larger values. The limit \(\lambda\rightarrow0\) for PMC is shown in gray dots, and the limit \(\lambda\rightarrow0\) for PEC coincides with the Yang-Mills Casimir energy \eqref{eq:energy-YM}, shown in red dots for reference. All numerical values are expressed in units \(\LambdaMS = 1\).}
    \label{fig:energy-PEMC}
\end{figure}

For easy comparison, we plot the GZ Casimir energy for PMC and PEC on the same plot in Fig.~\ref{fig:energy-PEMC}.

\subsubsection{Limit for \(\lambda\rightarrow0\)}\label{sec:lambda0}

As one can already anticipate from the different behavior of the GZ Casimir energy for PMC and PEC in Figs.~\ref{fig:energy-PMC}, resp.~\ref{fig:energy-PEC}, there exists a subtlety when taking the limit \(\lambda\rightarrow0\) (which is of course equivalent to \(\gamma\rightarrow0\)). One immediately sees that the non-local GZ action \eqref{eq:action-GZ-non-local} reduces to the FP action in this limit. So if one calculates the Casimir energy after letting \(\lambda\rightarrow0\), one gets that 
\begin{equation}\label{eq:energy-YM}
    \mathcal{E}_\text{Cas}^{\lambda\rightarrow0,\text{PMC}} = \mathcal{E}_\text{Cas}^{\lambda\rightarrow0,\text{PEC}} = \mathcal{E}_\text{Cas}^\text{YM}= - (N^2-1)\frac{\pi^2}{720 L^3},
\end{equation}
the well-known duality-invariant Yang-Mills Casimir energy. However, if one first calculates the Casimir energy in GZ theory, and only afterwards lets \(\lambda\rightarrow0\), one finds\footnote{
In the limit \(\lambda\rightarrow0\), we have that \(\bar \omega,\omega\rightarrow |\mathbf{k}|\), and thus \(x\rightarrow|\mathbf k|\) and \(y\rightarrow0\).
Since in the limit \(\lambda\rightarrow0\) the only scale left in the theory is \(L\), the discontinuity \(\mathcal{E}_\text{extra}\) necessarily behaves as \(1/L^3\), and can thus be expressed as a multiple of \(\mathcal{E}^\text{YM}_\text{Cas}\).
}
\begin{align}
    \lim_{\lambda\rightarrow0}\mathcal{E}_\text{Cas}^\text{PMC} &= \lim_{\lambda\rightarrow0} \mathcal{E}_1^\text{PMC} + 2 \lim_{\lambda\rightarrow0} \mathcal{E}_2^\text{PMC} \\* 
    &=  \frac{(N^2-1)}{4 \pi^2} \int_0^\infty dk\, k^2 \log \left[1 - \left( e^{-L k} \left(  1 + Lk \right) \right)^2\right] + 2\frac{(N^2-1)}{4 \pi^2} \int_0^\infty dk\, k^2 \log \left[1 - \left( e^{-L k} \left(  1 + 0 \right) \right)^2\right] \\*
    &= \mathcal{E}_\text{extra} + \mathcal{E}_\text{Cas}^\text{YM} \approx4.9 \times\mathcal{E}_\text{Cas}^\text{YM} , \label{eq:PMC-lim}\\
    \intertext{whereas}
    \lim_{\lambda\rightarrow0}\mathcal{E}_\text{Cas}^\text{PEC} &= 2\lim_{\lambda\rightarrow0} \mathcal{E}_2^\text{PEC} \\*
    &= 2\frac{(N^2-1)}{4 \pi^2}  \int_0^\infty dk\, k^2 \log \left[1 - \left( e^{-L k} \left(  1 - 0 \right) \right)^2\right] \\*
    &= \mathcal{E}_\text{Cas}^\text{YM}. \label{eq:PEC-lim}
\end{align}

We thus find that for PMC, the GZ Casimir energy is discontinuous upon taking the limit \(\lambda\rightarrow0\). More specifically, \(\mathcal{E}_2^\text{PMC}\) and \(\mathcal{E}_2^\text{PEC}\) approach each other in the limit, meeting at \(\mathcal{E}^\text{YM}_\text{Cas}\); but \(\mathcal{E}_1^\text{PMC}\) does not vanish when \(\lambda\rightarrow0\), causing the discontinuity. Informally, one can say that the residual gauge-fixing for PEC has eaten the energy contribution \(\mathcal{E}_1\). This is indeed what can be observed in the plots above.
This behavior is completely analogous to the Casimir energy in the Curci-Ferrari model, where one also finds this vDVZ-like discontinuity~\cite{vanDam:1970vg,Zakharov:1970cc} in the massless limit for PMC, and we refer the interested reader to~\cite{Dudal2026} for more details. 

\subsection{Casimir energy via the energy-momentum tensor}\label{sec:Casimir-EMT}

An alternative field theoretical approach for studying the vacuum energy is through the energy-momentum tensor (EMT). Computing the Casimir energy using a local energy density operator enables us to do an internal consistency check of the results found above. Moreover, this method is more flexible than the previous one, being more easily adapted for extensions such as non-planar or moving boundaries. Additionally, using the EMT approach, one can notice the possibility of having non-trivial boundary contributions to the EMT~\cite{Dudal:2024Robin,Dudal:2024PEMC,Dudal2026}, something that is usually not addressed in the literature.

In the following, we will obtain the canonical EMT by employing Noether's theorem for translation invariance:
\begin{equation}
    T_{\mu\nu} = \sum_{\text{fields }\phi}\frac{\delta \mathcal{L}}{\delta \partial_\mu \phi} \partial_\nu\phi - \delta_{\mu\nu}\mathcal{L}.
\end{equation}
The starting point is to discuss which Lagrangian we should use. The GZ action in its non-local form can lead to complications due to the presence of the inverse FP operator \(\mathcal{M}^{-1}\). Furthermore, since the EMT method for computing the Casimir energy is intrinsically local, it would be better to work in a local fashion from the beginning. Thus, in order to keep our reasoning as simple as possible, we will adopt the GZ Lagrangian in its local incarnation:
\begin{align}
\begin{split}
    \mathcal{L}_\text{GZ+BC} &= \frac{1}{4} F_{\mu \nu}^a F_{\mu \nu}^a + \frac{1}{2} H_{\mu \nu}^a F_{\mu \nu}^a - \frac{\alpha}{2} \kappa^a \kappa^a+ \kappa^a \partial_\mu A_\mu^a + \bar{c}^a \partial_\mu D_\mu^{ab} c^b \\*
    &- \bar{\varphi}^{ac}_\mu \mathcal{M}^{ab}\varphi^{bc}_\mu + \bar{\omega}^{ac}_\mu \mathcal{M}^{ab} \omega^{bc}_\mu  + \gamma^2 g f^{abc} A_\mu^b (\varphi_\mu^{bc} + \bar{\varphi}_\mu^{bc}) -4\gamma^4(N^2-1)
\end{split}
\end{align}

To be consistent with the approximations we have done so far, we will restrict ourselves to the quadratic level. Since we are only interested in $T_{00}$ (and will only use the diagonal part $T_{ii}$ in practice, see Eq.~\eqref{EMT00ii} below), we will not bother with symmetrization of the EMT. Furthermore, we perform partial integrations to rewrite the GZ action in a more suitable way, where each field is differentiated at most once. Integrating out the Nakanishi-Lautrup field for simplicity, we obtain:
\begin{align}
\begin{split}
    \mathcal{L}_\text{GZ+BC}^\text{quad} &= \frac{1}{4} (\partial_\mu A_\nu^a - \partial_\nu A_\mu^a)^2 +\frac{1}{2} H_{\mu \nu}^a (\partial_\mu A_\nu^a - \partial_\nu A_\mu^a) + \frac{1}{2\alpha} \left(\partial_\mu A_\mu^a\right)^2 - \partial_\nu \bar{c}^a \partial_\nu c^a  \\*
    &-\partial_\nu\bar{\varphi}_\mu^{ac} \partial_\nu \varphi_\mu^{ac} + \partial_\nu\bar{\omega}_\mu^{ac}  \partial_\nu \omega_\mu^{ac} + \gamma^2 g f^{abc} A_\mu^a \left(\varphi_\mu^{bc} + \bar{\varphi}_\mu^{bc} \right). 
\end{split}
\end{align}
To compute the contribution from the gauge field to the EMT, we calculate
\begin{align}
    \frac{\delta \mathcal{L}_\text{GZ+BC}^\text{quad}}{\delta \left(\partial_\mu A_\rho^b\right) }\partial_\nu A_\rho^b
    &= \left(\partial_\mu A_\rho^a - \partial_\rho A_\mu^a +H_{\mu\rho}^a \right) \partial_\nu A_\rho^a +  \frac{1}{\alpha} \partial_\alpha A_\alpha^a \partial_\nu A_\mu^a \\*
    &=\left(\partial_\mu A_\rho^a - \partial_\rho A_\mu^a +H_{\mu\rho}^a\right) \left(\partial_\nu A_\rho^a - \partial_\rho A_\nu^a \right)  + \frac{1}{\alpha} \partial_\mu A_\nu^a  \partial_\alpha A_\alpha^a + \frac{1}{\alpha} \partial_\nu  A_\mu^a \partial_\alpha A_\alpha^a + \gamma^2 g f^{abc} A_\nu^a \left(\varphi_\mu^{bc} + \bar{\varphi}_\mu^{bc}\right), 
\end{align}
where in the second line, we subtracted a total derivative \(\partial_\rho \left[\left(\partial_\mu A_\rho^a - \partial_\rho A_\mu^a +H_{\mu\rho}^a\right) A_\nu^a\right]\), and used the equation of motion for \(A\):
\begin{align}
    \partial_\rho \left(\partial_\mu A_\rho^a - \partial_\rho A_\mu^a +H_{\mu\rho}^a \right) =   \frac{1}{\alpha} \partial_\mu \partial_\alpha A_\alpha^a - \gamma^2 g f^{abc} \left(\varphi_\mu^{bc} + \bar{\varphi}_\mu^{bc}\right).
\end{align}
Next, we have to compute the contributions to the EMT from \(c,\bar c,\varphi,\bar \varphi,\omega,\bar \omega\). 
For \((c,\bar c)\), we find
\begin{align}
    \frac{\delta \mathcal{L}_\text{GZ+BC}^\text{quad}}{\delta \left(\partial_\mu c^a\right) }\partial_\nu c^a +  \frac{\delta \mathcal{L}_\text{GZ+BC}^\text{quad}}{\delta \left(\partial_\mu \bar{c}^a\right)} \partial_\nu \bar{c}^a 
    &= \partial_\mu \bar{c}^a \, \partial_\nu c^a + \partial_\nu \bar{c}^a \, \partial_\mu c^a.
\end{align}
Analogous calculations for \(\varphi\) and \(\omega\) yield contributions \(-(\partial_\mu \bar{\varphi}^{ac}_\rho \, \partial_\nu \varphi^{ac}_\rho + \partial_\nu \bar{\varphi}^{ac}_\rho \, \partial_\mu \varphi^{ac}_\rho)\), resp.~\(-(\partial_\mu \bar{\omega}^{ac}_\rho \, \partial_\nu \omega^{ac}_\rho + \partial_\nu \bar{\omega}^{ac}_\rho \, \partial_\mu \omega^{ac}_\rho)\).
Therefore, the EMT at the quadratic level is given by
\begin{align}
\begin{split}
   T_{\mu \nu}^\text{GZ+BC} &=  \left(\partial_\mu A_\rho^a - \partial_\rho A_\mu^a +H_{\mu\rho}^a\right) \left(\partial_\nu A_\rho^a - \partial_\rho A_\nu^a \right)  + \frac{1}{\alpha} \partial_\mu A_\nu^a  \partial_\alpha A_\alpha^a + \frac{1}{\alpha} \partial_\nu  A_\mu^a \partial_\alpha A_\alpha^a + \gamma^2 g f^{abc} A_\nu^a \left(\varphi_\mu^{bc} + \bar{\varphi}_\mu^{bc}\right)  \\*
   &+ (\partial_\mu \bar{c}^a  \partial_\nu c^a + \partial_\nu \bar{c}^a \partial_\mu c^a) - (\partial_\mu \bar{\varphi} \, \partial_\nu \varphi + \partial_\nu \bar{\varphi} \, \partial_\mu \varphi ) - (\partial_\mu \bar{\omega} \, \partial_\nu \omega + \partial_\nu \bar{\omega} \, \partial_\mu \omega ) \\*
   & -\delta_{\mu \nu}  \bigg[\frac{1}{4} (\partial_\alpha A_\beta^a - \partial_\beta A_\alpha^a)^2 +\frac{1}{2} H_{\alpha \beta}^a (\partial_\alpha A_\beta^a - \partial_\beta A_\alpha^a) + \frac{1}{2 \alpha} \left(\partial_\alpha A_\alpha^a\right)^2 -\partial_\alpha \bar{c}^a \partial_\alpha c^a  \\*
    & -\partial_\alpha\bar{\varphi}_\beta^{ac} \partial_\alpha \varphi_\beta^{ac} + \partial_\alpha\bar{\omega}_\beta^{ac}  \partial_\alpha \omega_\beta^{ac} + \gamma^2 g f^{abc} A_\alpha^a \left(\varphi_\alpha^{bc} + \bar{\varphi}_\alpha^{bc} \right) \bigg].
\end{split}
\end{align}

The Casimir energy can then be computed from the EMT using (see e.g.~\cite{Plunien:1986ca,Milton:2004ya,Bordag:1983zk})
\begin{equation}\label{T00int}
    \mathcal{E}_\text{Cas} = -\int_{-\infty}^{+\infty}dz \langle T_{00}^\text{GZ+BC} \rangle.
\end{equation}
As in \cite{Bordag:1983zk}, we integrate over the full $z$-axis, and not just over the interval between the plates. This is consistent with the fact that we did consider the quantum fields between and outside the plates, cf.~our path integral analysis with the auxiliary fields with the full $\mathbb{R}^4$ as bulk space.

In order to compute this expression \eqref{T00int}, we will first integrate out the localizing fields, and then the gauge field \(A\), such that only \(b\) remains. Then we can use the \(b\)-propagator, \(\mathbb{K}^{-1}\), to work out the VEV, cfr.~\cite[Sec.~4B]{Dudal:2024PEMC}.

We can integrate out all the fields but \(b\) using their equations of motion, listed here for convenience:
\begin{align}
    \partial_\rho \left(\partial_\mu A_\rho^a - \partial_\rho A_\mu^a +H_{\mu\rho}^a\right) &=   \frac{1}{\alpha} \partial_\mu \partial_\alpha A_\alpha^a - \gamma^2 g f^{abc} \left(\varphi_\mu^{bc} + \bar{\varphi}_\mu^{bc}\right),\\
    \partial^2c^a&=\partial^2\bar c^a =0,\\
    \varphi_\mu^{ab}&=\bar\varphi_\mu^{ab}=-\gamma^2 g f^{abc}\frac{1}{\partial^2}A_\mu^c,\\
    \partial^2\omega_\mu^{ab}&=\partial^2\bar\omega_\mu^{ab}=0.
\end{align}

We can immediately drop the decoupled \((c,\bar c)\) and \((\omega,\bar \omega)\), and fill in the equation of motion for \((\varphi,\bar \varphi)\). If we introduce the notation \(G_{\mu\rho}^a\) for the expression \(\partial_\mu A_\rho^a - \partial_\rho A_\mu^a\) considered on-shell, we get in Fourier space
\begin{equation}
    G_{\mu\nu}^a = \frac{k^2}{k^4+\lambda^4} k_\rho ( k_\mu H_{\nu\rho}^a - k_\nu H_{\mu\rho}^a),
\end{equation}
and the on-shell EMT as a function of \(b\) becomes
\begin{align}\label{eq:emt}
\begin{split}
    T^\text{GZ+BC}_{\mu\nu} &= (N^2-1)\bigg[ 
    G_{\mu\rho} G_{\nu\rho} - \frac14\delta_{\mu\nu} G_{\rho\sigma} G_{\rho\sigma}\\*
    &+H_{\mu\rho}G_{\nu\rho}-\frac12\delta_{\mu\nu} H_{\rho\sigma}G_{\rho\sigma}\\*
    &-\lambda^4 \frac{\partial_\mu}{\partial^2}A_\lambda[b] \frac{\partial_\nu}{\partial^2}A_\lambda[b] +\delta_{\mu\nu}\frac{\lambda^4}{2} \frac{\partial_\sigma}{\partial^2}A_\lambda[b] \frac{\partial_\sigma}{\partial^2}A_\lambda[b]\\*
    &-\lambda^4 A_\mu[b] \frac{1}{\partial^2}A_\nu[b] + \delta_{\mu\nu} \lambda^4 A_\lambda[b] \frac{1}{\partial^2}A_\lambda[b] \bigg],
\end{split}
\end{align}
where $N^2-1$ is a color global factor, the color indices are implicitly understood, and we use the notation \(A_\mu[b]\) for the gluon field written in terms of the boundary field by using its equation of motion. The cleanest form is found via the action~\eqref{eq:quad-action}, and in Fourier space it reads 
\begin{align}\label{AtobOnshell}
A_\mu[b] = i\frac{k^2}{k^4 + \lambda^4} k_\rho H_{\mu \rho}(k).    
\end{align}
In the EMT expression~\eqref{eq:emt}, one can easily identify the origin of the terms present in the different lines. Indeed,
the first line is the pure Yang-Mills EMT, the second line is the boundary contribution to the EMT, and the third and fourth line are the contribution from the localized horizon term~\eqref{eq:action-horizon-local}. For the upcoming calculations, we will use that 
\begin{equation}\label{EMT00ii}
	\mathcal{E}_\text{Cas} = -\int d{z} \left\langle T^\text{GZ+BC}_{00}\right \rangle =  -\frac13 \int d{z} \left\langle T^\text{GZ+BC}_{ii} \right\rangle.
\end{equation}
This expression is justified since in a Euclidean spacetime with plates along the \(z\)-direction, the other three directions are interchangeable (for a formal argumentation, see \cite[p.~9]{Dudal:2024PEMC}).

The EMT expression~\eqref{eq:emt} is written as a sum of contractions involving \(H_{\mu\nu}\), which can be written in Fourier space as \(H_{\mu\nu}(k)= e^{i k_z z^\gamma} H_{i \mu \nu} \, b_i^{\gamma}(\mathbf{k})\). To compute the Casimir energy, we have to compute $\left\langle T^\text{GZ+BC}_{ii} \right\rangle$, which in turn requires the computation of the boundary field propagator. Using Eq.~\eqref{bfprop}, the $b$-propagator can be written as
\begin{align}
    \langle b_l^{\gamma}(\mathbf{p}) b_k^{\lambda}(\mathbf{-p}) \rangle = (\mathbb{K}^{-1})^{\gamma\lambda}_{lk} = E^{\gamma\lambda}\Theta_{lk} +F^{\gamma\lambda} \Omega_{lk}.
\end{align}
The next step is to compute all the contractions involving \(H_{i \mu \nu}\). If one performs the algebra and works out all terms from Eq.~\eqref{eq:emt}, one encounters the following contractions for each boundary condition: 
\begin{align}
    \text{PMC:}\quad&\begin{cases}
        k_\sigma k_\beta H_{i\rho\sigma} H_{j\rho\beta} (\mathbb{K}^{-1})_{ij}^{\gamma\lambda} &\!\!\!\!\!= (k_z^2 \Theta_{ij} +k^2 \Omega_{ij})(\mathbb{K}^{-1})_{ij}^{\gamma\lambda}= 2k_z^2 E^{\gamma\lambda} + k^2 F^{\gamma\lambda},  \\
        k_\sigma k_\beta H_{ik\sigma} H_{jk\beta} (\mathbb{K}^{-1})_{ij}^{\gamma\lambda} &\!\!\!\!\!= (k_z^2 \Theta_{ij} +k_z^2 \Omega_{ij})(\mathbb{K}^{-1})_{ij}^{\gamma\lambda}= 2 k_z^2 E^{\gamma\lambda}  +k_z^2 F^{\gamma\lambda},\\
        k_k k_\beta H_{i\rho k} H_{j\rho\beta} (\mathbb{K}^{-1})_{ij}^{\gamma\lambda} &\!\!\!\!\!= \mathbf k^2 \Omega_{ij}(\mathbb{K}^{-1})_{ij}^{\gamma\lambda} \negphantom{\mathbf k^2 \Omega_{ij}(\mathbb{K}^{-1})_{ij}^{\gamma\lambda}} \phantom{(k_z^2 \Theta_{ij} +k_z^2 \Omega_{ij})(\mathbb{K}^{-1})_{ij}^{\gamma\lambda}}= \mathbf k^2 F^{\gamma\lambda}.
    \end{cases}\\*
    \text{PEC:}\quad &\phantom{\bigg\lbrace} k_\sigma k_\beta H_{i\rho\sigma} H_{j\rho\beta}(\mathbb{K}^{-1})_{ij}^{\gamma\lambda} \,\,= k_\sigma k_\beta H_{ik\sigma} H_{jk\beta}(\mathbb{K}^{-1})_{ij}^{\gamma\lambda} = k_k k_\beta H_{i\rho k} H_{j\rho\beta}(\mathbb{K}^{-1})_{ij}^{\gamma\lambda}  = -2\mathbf k^2 E^{\gamma\lambda}. 
\end{align}
With these expressions, the Casimir energy for PMC and PEC boundary conditions can be brought into the form
\begin{equation}
    \mathcal{E}_\text{Cas}=\int\!\frac{d^4k}{(2\pi)^4}\; e^{i k_z (z^\gamma-z^\lambda)} \left[  f(k)E^{\gamma\lambda} + g(k)F^{\gamma\lambda}\right],
\end{equation}
where $f(k)$ and $g(k)$ have lengthy expressions (that differ for the PMC and PEC cases) which will not be explicitly written here to avoid cluttering the text\footnote{The explicit Mathematica expressions can be obtained from the authors upon reasonable request.}. These expressions are naturally divergent and must be regularized. As we already discussed before, the physical Casimir energy is defined by the difference between the vacuum energy in the presence of boundaries and in their absence, which provides a natural way of regularizing these expressions. Remarkably, remembering that \(E^{\gamma \lambda}\) and \(F^{\gamma \lambda}\) only depend on \(\mathbf k\), one can perform the $dk_z$ integral analytically. However, we need to resort to numerical integration for the \(\mathbf k\)-integral. Comparing numerically the result for the Casimir energy obtained from the EMT approach with the analytical one obtained through the functional integral for both PMC~\eqref{eq:Cas-energy-PMC} and PEC~\eqref{eq:Cas-energy-PEC} boundary conditions, we find a perfect agreement, providing an extra verification of our results.


\section{Considerations on the gap equation}\label{sec:solving-gap-eq}

In the previous section, we computed the Casimir energy assuming that the Gribov parameter does not depend on $L$. 
However, as already emphasized, such a dependence could arise since the Gribov parameter is obtained by solving the gap equation. 
Indeed, it is reasonable to expect that such a solution can carry information about the plate separation: the gap equation is ultimately related to the vacuum energy, which depends on the location of non-trivial boundaries.
Thus, let us now have a closer look at the possible indirect $L$-dependence of $Z_A$ through $\lambda(L)$. 
The form of this dependence can be obtained by solving the gap equation while taking into account the modifications due to the non-trivial boundary. 
In order to do this, we first have to inspect the implications of the boundary-modified gluon propagator on the horizon condition, and then discuss the solutions of the gap equation under these circumstances.

\subsection{Boundary contribution to the gap equation}\label{sec:L-dependent-lambda}

The horizon condition~\eqref{eq:gap-eq-horizon} is given by \(\langle H(A) \rangle = 4 V_4 (N^2 - 1)\), where we emphasize that \(V_4\) denotes the four-dimensional volume. At the quadratic level, we have $\left[\mathcal{M}^{-1}_{xy}\right]^{ad} \approx \frac{\delta^{ad} \delta(x-y)}{(-\partial^2)}$. Thus, the horizon function~\eqref{eq:horizon-function} reads
\begin{equation}
    H(A) = Ng^2 \int d^4x A_\mu^a(x)\left(\frac{1}{-\partial^2}\right)A_\mu^a(x) = Ng^2 \int \frac{d^4p}{(2\pi)^d} \frac{1}{p^2}A_\mu^a(p)A_\mu^a(-p).
\end{equation}
The VEV in the horizon condition is to be taken with respect to the full action \(S\) in Eq.~\eqref{eq:action-full}, including the boundary term~\eqref{eq:SBC-2}. This means that for calculating \(\langle A_\mu A_\mu\rangle\), one needs to use the boundary-modified gluon propagator~\eqref{eq:A-prop}:
\begin{equation}
    \langle  A_\mu^a(p)A_\nu^b(q)\rangle =\delta^{(4)}(p+q) \left(K^{-1}\right)_{\mu \nu}^{ab} + D_{\mu\nu}^{ab}(p,q).
\end{equation}
With this propagator, we can work out the horizon condition to obtain the boundary-modified gap equation:
\begin{align}
    4 V_4 (N^2 - 1) &= \langle H(A) \rangle \\*
    &= Ng^2 \int \frac{d^4p}{(2\pi)^4} \frac{1}{p^2} \langle A_\mu^a(p)A_\mu^a(-p)\rangle \\*
    &= N g^2 V_4 \int \frac{d^4p}{(2\pi)^4} \frac{1}{p^2} (K^{-1})^{aa}_{\mu\mu} + N g^2 \int \frac{d^4p}{(2\pi)^4} \frac{1}{p^2} D_{\mu\mu}^{aa}(p,-p), \label{eq:gap-eq-bnd}
\end{align}
in which
\begin{align}
    (K^{-1})^{aa}_{\mu\mu} &= (N^2-1)3\frac{p^2}{p^4+\lambda^4},\\*
    D_{\mu\mu}^{aa}(p,-p) &= -V_3\frac{N^2-1}2 e^{-i p_z( z^\gamma - z^\lambda)} \left(\frac{p^2}{p^4+\lambda^4} \right)^2  p_\alpha H_{i\alpha\mu} \left(\mathbb{K}^{-1}\right)_{ij}^{\gamma\lambda} p_\beta H_{j\beta\mu}.
\end{align}
If we only had the first term in \(K^{-1}\), we would get the usual gap equation. But the boundary term in \(D_{\mu\mu}^{aa}\) is new and might alter the gap equation, potentially introducing \(L\)-dependence, which would result in an implicit dependence of the Gribov mass on \(L\). We first note that the second  integral in \eqref{eq:gap-eq-bnd} is power counting convergent, since in the UV it goes like $\int dp/p^2$. Now, since the boundary term only carries a 3-dimensional volume factor \(V_3\), the boundary term will be dominated by the 4D term. Or in other words, if we divide the gap equation by \(V_4\), the boundary term is proportional to \(V_3/V_4 = 1/\ell_z =0 \) because of the infinite length of spacetime in the \(z\)-direction. We can thus safely conclude that the presence of plates in the system does not modify the gap equation, and therefore it was justified to treat the Gribov parameter as being $L$-independent in Section.~\ref{sec:free-Gribov-param}.


\subsection{Casimir energy for a specific solution of the gap equation}\label{sec:MSbar-solution}

The gap equation has been solved in empty space in \cite[Eq.~(133-137)]{Dudal:2008sp}. Let us briefly revisit this solution here. The gap equation in the $\msbar$ renormalization scheme is given by
\begin{align}\label{eq:gap-eq-to-solve}
\frac{\partial\mathcal{E}}{\partial\lambda} = 4\lambda^3\left( -\frac{2(N^2-1)}{g^2 N} + \frac{3(N^2-1)}{64\pi^2} \left(\frac53 -2\ln \frac{\lambda^2}{\mu^2}\right)\right) = 0.
\end{align}
Looking at Eq.~\eqref{eq:gap-eq-to-solve}, it is convenient to choose $\mu = \lambda$ such that the logarithm in the gap equation vanishes. This effectively corresponds to a resummation of the (leading) logs into the coupling constant. In this case, ignoring the trivial case $(\lambda=0)$, one immediately finds the following solution for the gap equation: 
\begin{align}\label{solutiongap2}
 \frac{N g^2(\lambda)}{16 \pi^2} = \frac{8}{5}. 
\end{align}
Plugging this solution for the gap equation into the one-loop running coupling expression, 
\begin{align}\label{RunningCoupling}
   g^2(\mu) = \frac{1} {\beta_0\log\left(\frac{\mu^2}{\Lambda^2}\right)},\qquad \beta_0=\frac{11}{3}\frac{N}{16\pi^2},
\end{align}
where $\Lambda$ is the renormalization group invariant scale of the considered scheme, one finally gets in the $\msbar$ scheme \footnote{Note the typo in \cite{Dudal:2008sp}: Eq.~(137) should read $\lambda^4 = e^{15/44} \approx 1.41$ instead of $\lambda^4 = e^{44/15}$, in units $\lms=1$.}
\begin{align} \label{lambdaFromLambda}
    \lambda = e^{15/176} \lms \approx 1.09 \, \lms.
\end{align}
A major drawback of this solution for $\mu=\lambda$ is that the perturbative expansion is not reliable, since the expansion parameter \eqref{solutiongap2} is too large. 

\begin{figure}
    \centering
    \includegraphics[width=0.6\linewidth]{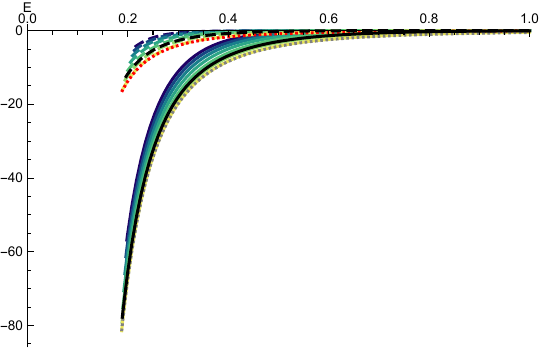}
    \caption{Identical legend as Fig.~\ref{fig:energy-PEMC}, with the $V$-scheme solution $\lambda = e^{\frac{541}{528}}$ printed on top in black.}
    \label{fig:energy-PEMC-Vscheme}
\end{figure}

Let us henceforth look for a more suitable renormalization scheme. We recall here that the renormalization of the GZ theory, defined via its action~\eqref{eq:action-GZ} only requires 2 independent renormalization factors, see e.g.~\cite{Zwanziger93,Dudal:2008sp}. We make the choice to still renormalize $\lambda$ itself in the $\msbar$ scheme, while for the strong coupling constant, we will adopt the so-called $V$-scheme. This corresponds to imposing that the (perturbative\footnote{For completeness, we note there that the static potential was computed in \cite{Gracey:2009mj} also in presence of the Gribov mass $\lambda$, but generalizing the $V$-scheme to this setting leads to an utterly complicated, and explicitly $\lambda$-dependent scheme. Our strategy is the one of \cite{Zwanziger93}: we can maintain the renormalization factors as stemming from the standard Yang-Mills action (corresponding to $\lambda\to0$), as afterwards all quantities computed from the GZ action are also properly renormalized.}) static Coulomb potential $V(q^2)$ between two infinitely heavy test charges remains given by its classical value, upon replacement of the coupling constant with $g^2_V(q^2)$, thereby defining the $V$-scheme effective coupling constant, see
\cite{Peter:1996ig} for more:
\[V(q^2)=-\frac{C_2(R) g^2_V(q^2)}{q^2},\qquad \text{Tr}(t^a t^a)=C_2(R)\mathds{1}_R\]
where $t^a$ are the generators of the $R$-representation. At one-loop order, the correspondence between this coupling and its $\msbar$ counterpart $g^2$ reads \cite[eq.(11)]{Peter:1996ig}
\[g^2_V=g^2(1-b_0g^2),\qquad b_0=-\frac{1}{16\pi^2}\frac{31N}{9}\]
Implementing this transformation at the level of the gap equation yields\begin{align}\label{eq:gap-eq-to-solve2} 
 -\frac{2(N^2-1)}{g^2_V N}\left(1-\frac{31}{9}\frac{g_V^2N}{16\pi^2}\right) + \frac{3(N^2-1)}{64\pi^2} \left(\frac53 -2\ln \frac{\lambda^2}{\mu^2}\right) = 0.
\end{align}
Setting once more $\mu=\lambda$ to resum the logs, we now find
\begin{align}\label{solutiongap3}
 \frac{g^2(\lambda)N}{16 \pi^2} = \frac{72}{293} \approx 0.25
\end{align}
which is already far more acceptable than  \eqref{solutiongap2}, so that we can trust to some extent the ensuing perturbative estimate of the Gribov mass. The standard conversion rule from \cite{Celmaster:1979km}, $\Lambda_V=\lms e^{-\frac{b_0}{2\beta_0}}$, leads to
\begin{align} \label{lambdaFromLambdabis}
    \lambda = e^{\frac{293}{528}} \Lambda_V=e^{\frac{293}{528}} e^{-\frac{b_0}{2\beta_0}} \lms = e^{\frac{541}{528}}\lms \approx  2.79\lms.
\end{align}
We plot the Casimir energy for this solution \eqref{lambdaFromLambdabis} in Fig.~\ref{fig:energy-PEMC-Vscheme}.

For the record, one might question the usage of the usual running coupling expression \eqref{RunningCoupling} as we did alter the underlying quantum field theory. Said otherwise, could the renormalization factor for $g^2$ change due the planar interfaces? In \cite{Symanzik:1981wd}, it was discussed that the usual bulk counterterms suffice, in combination with suitable surface counterterms, per surface, to render the considered theories fully renormalizable. In any case, the bulk running coupling is unaffected and certainly does not require a location dependent renormalization. A rigorous proof for the case of $\lambda\phi^4$ on a half-space can be found in \cite{Borji:2022kuw,Borji:2023eyp}. Intuitively, this is understandable, as UV divergences come about from quantum fields ``touching'' upon each other in spacetime. Either this happens in the bulk and then it has nothing to do with interfaces, or on an interface itself.

\subsection{Comparing with lattice data}
Given the success of recent lattice simulations~\cite{Chernodub:2023dok, Chernodub:2018pmt, Ngwenya:2025cuw,Ngwenya:2025mpo, Chernodub:2019nct}, we have several Casimir energy datasets at our disposal for comparing our analytical results against.
Our analytical expression~\eqref{eq:Cas-energy-PEC} for the PEC Casimir energy is able to fit the relevant lattice data, provided that one includes a (customary) phenomenological global prefactor $C$, see \cite{Chernodub:2023dok}.

\begin{figure}[t!]
	\begin{minipage}[b]{0.5\linewidth}
\includegraphics[width=\textwidth]{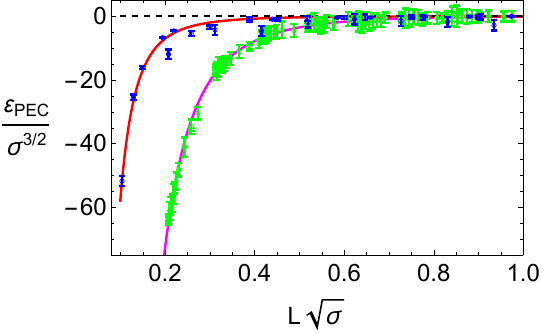}
	\end{minipage} \hfill
\caption{Non-Abelian Casimir energy between parallel plates under PEC boundary conditions for $SU(3)$ gauge group in 4D in units of $\sqrt{\sigma}$. The red and magenta curves represent the analytical expression~\eqref{eq:Cas-energy-PEC} fitted on the lattice data of~\cite{Chernodub:2023dok} (in green) and~\cite{Ngwenya:2025cuw} (in blue), respectively.}
\label{CasimirGZ4dSU3Joint}
\end{figure}

\begin{figure}[t!]
	\begin{minipage}[b]{0.5\linewidth}
\includegraphics[width=\textwidth]{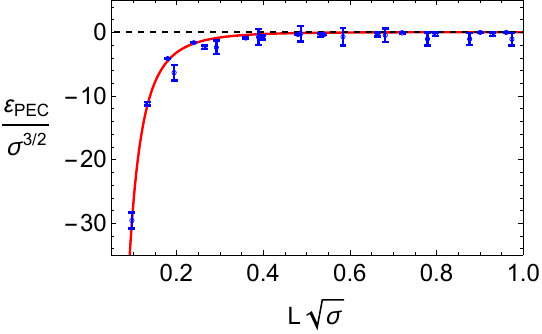}
	\end{minipage} \hfill
\caption{Non-Abelian Casimir energy between parallel plates under PEC boundary conditions for $SU(2)$ gauge group in 4D in units of $\sqrt{\sigma}$. The red curve represents the analytical expression~\eqref{eq:Cas-energy-PEC} fitted on the lattice data of~\cite{Ngwenya:2025cuw} (in blue).}
\label{CasimirGZ4dSU2}
\end{figure}

In 4D, we can immediately use the PEC Casimir energy \eqref{eq:Cas-energy-PEC}. There are two manners in which one can make the fit. The first one is to do a dual fit: fit both a global multiplicative factor $C$ and a value for $\lambda$ at the same time. The second manner is to do a single fit for the global multiplicative factor $C$, and use the $V$-scheme value for $\lambda$ as obtained above in \eqref{lambdaFromLambdabis}. The expression \eqref{lambdaFromLambdabis} should be reexpressed in terms of the string tension $\sigma$, so in order to get a numerical value, we need a numerical estimate for $\LambdaMS/\sqrt\sigma$. We use the estimates $\LambdaMS/\sqrt\sigma = 0.538$ for $SU(3)$, and $\LambdaMS/\sqrt\sigma= 0.752$ for $SU(2)$, see \cite{Lucini:2008vi}. We thus get $\lambda^{SU(3)}_{V} = e^{\frac{541}{528}} \times 0.538 \approx 1.50$, and $\lambda^{SU(2)}_V = e^{\frac{541}{528}} \times 0.752 \approx 2.10$, from now on working in units $\sqrt\sigma=1$.
Both fitting procedures match the data very well, which we show in Fig.~\ref{CasimirGZ4dSU2} for $SU(2)$ and Fig.~\ref{CasimirGZ4dSU3Joint} for $SU(3)$. The fitted values can be found in Table \ref{tab:4d}.

\begin{table}
    \centering
    \begin{tabular}{c|cc|cc|}
        4D & \multicolumn{2}{c|}{Single fit} & \multicolumn{2}{c}{Dual fit} \\ \hline
        $SU(3)$ \cite{Chernodub:2023dok} & $\lambda^{SU(3)}_{V} = 1.50$ & $C=5.48$ & $\lambda_\text{fit}=1.10$ & $C=5.60$ \\
        $SU(3)$ \cite{Ngwenya:2025cuw} & $\lambda^{SU(3)}_{V} = 1.50$ & $C=0.53$ & $\lambda_\text{fit} = 0.015$ & $C=0.53$ \\ 
        $SU(2)$ \cite{Ngwenya:2025cuw} & $\lambda^{SU(2)}_{V} = 2.10$ & $C=0.68$ & $\lambda_\text{fit} = 0.018$ & $C=0.67$\\ \hline
    \end{tabular}
    \caption{Parameter values for which the PEC Casimir energy \eqref{eq:Cas-energy-PEC} best fits the available lattice data in 4D.}
    \label{tab:4d}
\end{table}

\begin{figure}[t!]
\centering
\begin{subfigure}{.5\textwidth}
  \centering
  \includegraphics[width=\linewidth]{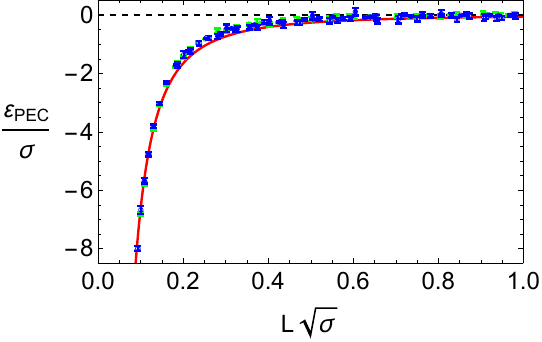}
  \label{fig:su2-sol}
\end{subfigure}%
\begin{subfigure}{.5\textwidth}
  \centering
  \includegraphics[width=\linewidth]{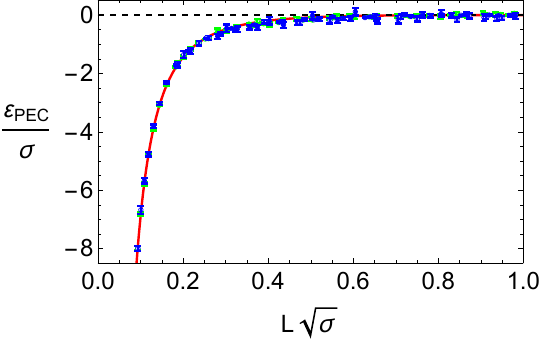}
  \label{fig:su2-fit}
\end{subfigure}
\caption{Non-Abelian Casimir energy between parallel wires under PEC boundary conditions for $SU(2)$ gauge group in 3D in units of $\sqrt{\sigma}$. The red curve represents the analytical expression~\eqref{eq:Cas-energy-PEC} fitted to the lattice data of \cite{Chernodub:2018pmt} (in green) and \cite{Ngwenya:2025mpo} (in blue) (both fits overlap). Left, the leading order gap solution $\lambda^{SU(2)}_\text{gap} = 0.22$ is used; right, $\lambda$ is fitted on the data.}
\label{CasimirGZ3dSU2Joint}
\end{figure}
\begin{figure}[t!]
\centering
\begin{subfigure}{.5\textwidth}
  \centering
  \includegraphics[width=\linewidth]{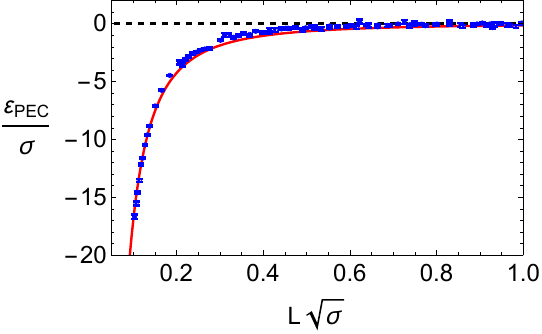}
  \label{fig:su3-sol}
\end{subfigure}%
\begin{subfigure}{.5\textwidth}
  \centering
  \includegraphics[width=\linewidth]{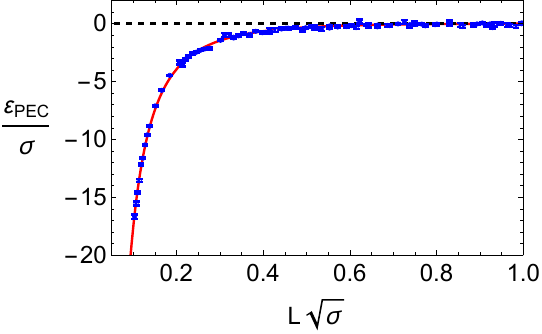}
  \label{fig:su3-fit}
\end{subfigure}
\caption{Non-Abelian Casimir energy between parallel wires under PEC boundary conditions for $SU(3)$ gauge group in 3D in units of $\sqrt{\sigma}$. The red curve represents the analytical expression~\eqref{eq:Cas-energy-PEC} fitted on the lattice data of \cite{Ngwenya:2025mpo} (in blue). Left, the leading order gap solution $\lambda^{SU(3)}_\text{gap} = 0.20$ is used; right, $\lambda$ is fitted on the data.}
\label{CasimirGZ3dSU3}
\end{figure}

We can repeat this comparison with lattice data in 3D, since our analytical expression \eqref{eq:Cas-energy-PEC} for the 4D PEC Casimir energy can readily be modified to yield the 3D PEC Casimir energy between two wires:
\begin{equation}\label{eq:Cas-energy-PEC-3d}
    \mathcal{E}_\text{Cas wire}^\text{PEC} = \frac{(N^2-1)}{4 \pi} \int_0^\infty dk\, k \log \left[1 - \frac{(\bar \omega e^{-L \omega} + \omega e^{-L \bar \omega} )^2}{(\bar \omega + \omega)^2} \right].
\end{equation}
However, in 3D the gap equation is different and yields a finite leading-order solution $\lambda^{SU(N)}_\text{gap} = \frac{\sqrt{2}}{12\pi} g^2 N$, cfr.~\cite[Eq.~(24)]{Dudal:2008rm}.\footnote{We refer to the discussion in Section \ref{sec:no-pole-condition} for why the presence of the plates does not modify the gap equation in 3D either.} The coupling constant $g^2$ is dimensionful now and is related to the string tension $\sigma$ via $\frac{\sqrt{\sigma}}{g^2 N} = 0.20 - \frac{0.12}{N^2}$, cfr.~\cite[Eq.~(9)]{Lucini:2002wg}. Combining both equations yields numerical solutions for the 3D gap equation $\lambda^{SU(2)}_\text{gap} = 0.22$ and $\lambda^{SU(3)}_\text{gap} = 0.20$ in units of $\sqrt{\sigma}$.
Having these values, one can choose to perform a dual fit (of both $C$ and $\lambda$) or a single fit (only fit $C$ and use the gap solution $\lambda^{SU(N)}_\text{gap}$). We show the result in Fig.~\ref{CasimirGZ3dSU2Joint} for $SU(2)$, in Fig.~\ref{CasimirGZ3dSU3} for $SU(3)$, and the fitted values in Table \ref{tab:3d}.

\begin{table}
    \centering
    \begin{tabular}{c|cc|cc|}
        3D & \multicolumn{2}{c|}{Single fit} & \multicolumn{2}{c}{Dual fit} \\ \hline
        $SU(3)$ \cite{Ngwenya:2025cuw} & $\lambda^{SU(3)}_\text{gap} = 0.20$ & $C=0.90$ & $\lambda_\text{fit}=1.65$ & $C=0.97$ \\
        $SU(2)$ \cite{Chernodub:2023dok} & $\lambda^{SU(2)}_\text{gap} = 0.22$ & $C=0.93$ & $\lambda_\text{fit} = 1.86$ & $C=1.01$\\ 
        $SU(2)$ \cite{Ngwenya:2025cuw} & $\lambda^{SU(2)}_\text{gap} = 0.22$ & $C=0.92$ & $\lambda_\text{fit} = 1.70$ & $C=0.99$ \\ \hline
    \end{tabular}
    \caption{Parameter values for which the PEC Casimir energy \eqref{eq:Cas-energy-PEC} best fits the available lattice data in 3D.}
    \label{tab:3d}
\end{table}

Let us briefly discuss these results. Comparing the dual fit and single fit scheme, a first observation is that the fit of the global prefactor $C$ is not very sensitive to the value of $\lambda$. This is completely analogous to the situation for the Curci-Ferrari model: there, the fits were not very sensitive to the mass value, see \cite{Dudal2026}.
It is also worth noting that the values of $C$ we find here are almost identical to the values found for the Curci-Ferrari case in \cite{Dudal2026}.

As was noted before in \cite{Dudal2026}, in 3D, the fitted values for the prefactor are close to one, $C\approx1$, which indicates that most of the Casimir physics is already captured by the GZ model at leading order. 

In the 4D case, as discussed in~\cite{Dudal2026,Karabali:2025olx}, there is a rather large mismatch between the SU(2) data sets of~\cite{Chernodub:2018pmt} and~\cite{Ngwenya:2025cuw}, and consequently, in the fitted $C$-factors. A precise understanding of the  physical mechanism underlying the origin of this $C$ is lacking as of now. If $C\approx 1$, it is reasonable to assume it is related to the neglected (presumably small) loop corrections to the Casimir energy, but larger $C$'s cannot be explained in such a way. We will sketch a few possible sources of missing effects in the next section, which will require further exploration in the future.


\section{No-pole condition for the ghost propagator with broken translation invariance}\label{sec:no-pole-condition}
As mentioned before, the restriction to the Gribov region is usually achieved through a no-pole condition, which relies on the intrinsic relationship between the ghost propagator and the FP operator. However, the construction of \cite{Gribov78,Capri13,Zwanziger89} assumes full translation invariance of the system, which is partially broken in the present scenario, due to the presence of the parallel plates. Therefore, it is crucial to investigate the consequences of a partially broken translation symmetry in the implementation of the no-pole condition, for it might change the horizon function, and thus the action \eqref{eq:action-GZ-non-local} we are considering. As this turns out to be a rather tough problem, in this paper we will limit ourselves to a leading order analysis, and this only for PEC boundary condition. We hope to come back to the more challenging PMC case in later work.

Let us now start the discussion. In settings with translation invariance, the FP operator is diagonal in momentum space, namely $\mathcal M^{ab} \propto \delta^{(d)}(k-q)$. This means that its inverse, the ghost propagator, is as well:
\begin{equation}
	\langle \bar{c}_a(k)c_b(q)\rangle = \delta^{ab} \delta^{(d)}(k-q) F(k) \;.
\end{equation}
Positivity of the FP operator is then equivalent with having no poles in the ghost propagator, except at vanishing momentum.

Our case is more complicated, though. Due to the absence of momentum conservation, the FP operator is a non-diagonal operator $\mathcal M^{ab}(k,q)$. One could in theory again rewrite the ghost propagator in a 1PI resummation, but the result would be the inverse of some operator. One should then demand that this utterly complicated operator has no zero modes, which is far from trivial.

Instead, we shall proceed like in \cite{Zwanziger89} and compute the eigenvalues of the FP operator in a perturbative series in the gluon fields. The restriction to the Gribov region can then be imposed by demanding that those eigenvalues remain positive.

To do so, split the FP operator:
\begin{equation}\label{fpsplit}
	\mathcal M^{ab} = \mathcal M^{ab}_0 + \mathcal M^{ab}_1 \qquad \text{with} \qquad \mathcal M^{ab}_0 = - \delta^{ab} \partial^2 \;, \qquad \mathcal M^{ab}_1 = gf^{abc}  A_\mu^c \partial_\mu \;.
\end{equation}
The operator $\mathcal M^{ab}_0$ is diagonal in momentum space and will be taken as the leading order; $\mathcal M^{ab}_1$ is the perturbation. The operator $\mathcal M^{ab}_0$ has many identical eigenvalues: all plane waves with same momentum-squared have the same eigenvalue. We will therefore need to use degenerate perturbation theory. The problem is simplified by the observation that the risk of negative eigenvalues after perturbation is biggest for small momentum, such that we will be interested in the limit of zero momentum. In other words, we are interested in infinitesimal (though non-vanishing) momenta.

Just as when computing the ghost propagator for the no-pole condition in Gribov's original approach~\cite{Gribov78}, one needs to take into account the later integration over the gluon fields. This means that the first term in the perturbative series of the eigenvalues of the FP operator is
\begin{equation}
	\langle k | \mathcal M_1 | k \rangle \propto \langle k | \tilde A_\mu | k \rangle \qquad \rightarrow \qquad 0 \;.
\end{equation}
We temporarily ignored the color structure of the perturbation, and the bra-kets $\langle k|$ and $|k\rangle$ denote plane waves. The expectation value of $\tilde A_\mu$ is zero, which is why this matrix element drops out of consideration.

For the second term in the perturbation series, we will need expressions of the form $\langle k|\mathcal M_1|p\rangle\langle p|\mathcal M_1|q\rangle$. The momenta $k$ and $q$ need not be the same, as in degenerate perturbation it is often necessary to change the basis. This means we need to compute expressions involving the expectation value of $k_\mu p_\nu\tilde A_\mu^a(p-k)\tilde A_\nu^b(q-p)$. After integrating out the gluon fields, we still have invariance under color rotations, spatial rotations around the $z$ axis and translations perpendicular to the $z$ axis. The most general form of $k_\mu p_\nu\langle\tilde A_\mu^a(p-k)\tilde A_\nu^b(q-p)\rangle$ consistent with these constraints is computed in \appendixname\ \ref{appgluon}.

In principle we should now compute the perturbative shift to any given eigenvalue of $\mathcal M^{ab}_0$, but just as in the usual case, we are going to limit ourselves to several well-chosen eigenvalues. As in the standard case, we focus on infinitesimal momenta, since these are the most susceptible to becoming negative once perturbative corrections are taken into account. There are two main limits we can take when the momentum becomes small: we can take $k_z=0$ first and $\mathbf k\to0$ after, or $\mathbf k=\mathbf0$ first and $k_z\to0$ after.

The perturbation respects these limits: an eigenvector of $\mathcal M_0^{ab}$ with $k_z=0$ does not mix up with those with $k_z\neq0$, and the same for $\mathbf k$. Furthermore, the perturbation does not modify $\mathbf k$. It can, however, mess with $k_z$: If we start from a state with certain $k_z$, it can mix with the state with opposite sign for $k_z$, as those have the same eigenvalue for $\mathcal M_0^{ab}$ and there is no conservation rule prohibiting this mixing due to breaking of translation invariance in the $z$ direction. 

First, let us look at the case with $k_z=q_z=0$ in \eqref{perturb}. This gives the perturbed eigenvalue
\begin{multline}
	\mathbf k^2\Bigg(1 - \frac{g^2N}{(N^2-1)(d-2)} \int \frac{d^dp}{(2\pi)^d} \frac1{p^2+2\mathbf p\cdot\mathbf k}\Bigg(\left(1-\frac{(\mathbf k\cdot\mathbf p)^2}{\mathbf k^2\mathbf p^2}\right) \tilde A_i^c(p)\tilde A_i^c(-p) \\
	+ \left(1-(d-1)\frac{(\mathbf k\cdot\mathbf p)^2}{\mathbf k^2\mathbf p^2}\right) \frac{p_z^2}{\mathbf p^2} \tilde A_z^c(p)\tilde A_z^c(-p)\Bigg) + \cdots\Bigg) \;,
\end{multline}
where we shifted $\mathbf p\to\mathbf p+\mathbf k$ in the integration. Just like in the usual case, we can take the limit $\mathbf k\to\mathbf 0$. To do so, expand in small $\mathbf k$ and take $k_ik_j \to \frac1{d-1}\mathbf k^2\delta^{ij}$. This yields the Gribov condition
\begin{equation} \label{eq:condv}
	 0 < 1 - \frac{g^2N}{(N^2-1)(d-1)} \int \frac{d^dp}{(2\pi)^d} \frac1{p^2} \tilde A_i^c(p)\tilde A_i^c(-p) + \cdots \;.
\end{equation}

For the case with $\mathbf k=0$, we have to consider the two states with $\pm k_z$. It turns out that the perturbation lifts the degeneracy, and the new eigenstates are the symmetric and antisymmetric combinations of the states with $\pm k_z$. The perturbed eigenvalues are
\begin{multline}
	k_z^2 \Bigg( 1 - \tfrac12 \frac{g^2N}{N^2-1} \int \frac{d^dp}{(2\pi)^d} \left(\frac1{p^2+2p_zk_z} \tilde A_z^c(p)\tilde A_z^c(-p) \pm \frac1{p^2+2p_zk_z} \tilde A_z^c(p)\tilde A_z^c(-\mathbf p,-p_z-2k_z) \right. \\
	\left. \pm \frac1{p^2-2p_zk_z} \tfrac12\tilde A_z^c(p)\tilde A_z^c(-\mathbf p,-p_z+2k_z) + \frac1{p^2-2p_zk_z} \tilde A_z^c(p)\tilde A_z^c(-p)\right) + \cdots \Bigg) \;,
\end{multline}
where we once more shifted $p_z\to p_z\pm k_z$, depending on the term. The two terms with violation of momentum conservation are due to the mixing of unperturbed eigenstates with different sign for $k_z$. Taking $p\to-p$ in the last two terms whilst again keeping the exchange symmetry stated below \eqref{exchange} in mind, shows they are equal to the first two terms, and we obtain two Gribov conditions:
\begin{equation} \label{eq:condpm}
	0 < 1 - \frac{g^2N}{N^2-1} \int \frac{d^dp}{(2\pi)^d} \frac1{p^2+2p_zk_z} \left(\tilde A_z^c(p)\tilde A_z^c(-p) \pm \tilde A_z^c(p)\tilde A_z^c(-\mathbf p,-p_z-2k_z) \right) + \cdots \;.
\end{equation}
We will not yet take $k_z\to0$ at this point, in order to correctly deal with the violation of momentum conservation.

Now we introduce these conditions into the path integral through the Fourier representation of the Heaviside function:
\begin{equation}\label{heavi}
	\int_{-i\infty+\varepsilon}^{+i\infty+\varepsilon} \frac{d\beta_{x}}{2\pi i\beta_{x}} e^{\beta_x\text{condition}_x} \;,
\end{equation}
where the index $x$ enumerates the conditions. For the right-hand side of \eqref{eq:condv} we introduce a $\beta_{\text{v}}$, and for the right-hand sides of \eqref{eq:condpm} we introduce $\beta_\pm$. Putting $\beta_++\beta_- = \beta_z$ and $\beta_+-\beta_- = \beta_{\text{br}}$, we get as extra contribution to the action
\begin{multline} \label{eq:threeGribov}
	\ln\beta_{\text{v}} - \beta_{\text{v}} + \beta_{\text{v}}\frac{g^2N}{(N^2-1)(d-1)} \int \frac{d^dp}{(2\pi)^d} \frac1{p^2} \tilde A_i^a(p)\tilde A_i^a(-p) \\
	+ \ln(\beta_z^2-\beta_{\text{br}}^2) - \beta_z + \beta_z \frac{g^2N}{N^2-1} \int \frac{d^dp}{(2\pi)^d} \frac1{p^2+2p_zk_z} \tilde A_z^a(p)\tilde A_z^a(-p) \\
	+ \beta_{\text{br}} \frac{g^2N}{N^2-1} \int \frac{d^dp}{(2\pi)^d} \frac1{p^2+2p_zk_z} \tilde A_z^a(p)\tilde A_z^a(-\mathbf p,-p_z-2k_z) \;.
\end{multline}
These additions modify the expression for $K^{-1}$ in \eqref{eq:K-inv}. This needs to be computed before we can add the plates. First put
\begin{equation}
	\gamma_{\text{v}}^4 = \beta_{\text{v}} \frac2{V^{(d)}} \frac{g^2N}{(N^2-1)(d-1)} \;, \qquad \gamma_{z,\text{br}}^4 = \beta_{z,\text{br}} \frac2{V^{(d)}} \frac{g^2N}{N^2-1} \;.
\end{equation}
Then, we find in \appendixname\ \ref{app:kinv} that $(K^{-1})^{ab}_{\mu\nu}(p,q)$ equals $\delta^{ab}\delta^{(d-1)}(\mathbf p-\mathbf q)$ times the expression \eqref{eq:kwinv}.

Another expression we need is the operator $\mathbb K$ in \eqref{eq:Sb-quad}. As announced before, from now on we will solely consider the PEC case. The operator $\mathbb K$ is computed from \eqref{eq:Sb}, namely by contracting $K^{-1}$ with $p_\nu H_{\nu\mu}^a(p) = ip_j e^{i p_z z^\gamma} b_i^{\gamma, a}(\mathbf p) \varepsilon_{ij\mu z}$. This gives zero when multiplying with the second term in \eqref{eq:kwinv}, and so the only change to \eqref{eq:KK} is that $\lambda$ is replaced with $\lambda_{\text{v}}$. In the PMC case, this nullification does not happen and more ``annoying terms'' need to be explicitly dealt with.

The next step is to write down the steepest-descent equations. To do this, write down the generating functional without sources. One readily finds
\begin{equation}
	\frac1{\beta_{\text{v}}(\beta_z^2-\beta_{\text{br}}^2)} e^{\beta_{\text{v}}+\beta_z} [\det K]^{-1/2} [\det \mathbb K]^{-1/2} \;.
\end{equation}

Now, the presence of the plates is only visible through the operator $\mathbb K$, which only depends on $\beta_{\text{v}}$. This means that the gap equations for $\beta_z$ and $\beta_{\text{br}}$ do not depend on the presence of the plates. In the absence of plates, there can be no violation of momentum conservation, such that the gap equation for $\beta_{\text{br}}$ must be solved with $\beta_{\text{br}} = 0$.

As a result, we can also safely take the limit $k_z\to0$ at this point. We could not do so earlier, as explained in \appendixname\ \ref{app:kinv}.

Then, when writing down the gap equation for $\beta_{\text v}$, we need $\frac{V^{(d)}}2 \operatorname{tr}_d \left(\frac{dK}{d\beta_{\text{v}}} K^{-1}\right)$ and $\frac{V^{(d-1)}}2 \operatorname{tr}_{d-1} \left(\frac{d\mathbb K}{d\beta_{\text{v}}} \mathbb K^{-1}\right)$. The volume factors were added because we are taking the trace of operators with continuous spectrum: $\sum_\lambda \to V \int d\lambda$. The operator $\mathbb K$ lives in one dimension less that $K$, such that the $\mathbb K$ term drops out in the infinite-volume limit.

As such, all influence of the plates drops out of the gap equations. The solution must therefore be equal to the case without plates, which also means that $\lambda_{\text{v}} = \lambda_z$, due to the rotational invariance in that case. Said otherwise, we do end up with the original GZ action, in line with Sec.~\ref{sec:GZ}. Although this is open for further study, we expect that the extra PMC terms mentioned before become, in the infinite volume limit, obsolete as well when compared to the leading 4D bulk piece.


\section{Conclusion and outlook}\label{sec:conclusion}

We have studied Gribov-Zwanziger theory in 4D and 3D with planar boundaries using functional integral methods. 

We considered the specific boundary setup consisting of two infinite parallel plates, separated by a distance \(L\), and applied perfect magnetic or perfect electric boundary conditions to the plates. 
By lifting these boundary conditions into the action using Lagrange multiplier fields, we could compute the gluon propagator affected by these PMC resp.~PEC boundaries.
Since this modified gluon propagator feels the distance between the plates, one might expect that this \(L\)-dependence is transferred to the horizon condition \eqref{eq:gap-eq-horizon}. 
Because the Gribov mass \(\lambda\) solves the horizon condition dynamically, this would create an indirect dependence of the Gribov mass on \(L\). 
As the gap equation is equivalent to extremizing the vacuum energy, this would mean that the Casimir energy might get a correction via the \(L\)-dependent Gribov mass. 
However, we showed that no such dynamical \(L\)-dependence of the Gribov mass occurs because of the mismatching of volume factors. 

A priori, the breaking of translation symmetry might invalidate the usual argument that restricts the integration area to the Gribov region via a horizon term in the action. 
However, we gave plausible evidence that this argument also works without asserting translation invariance, leaving the horizon term unchanged, at least if the bulk space is assumed to be of infinite length in the direction orthogonal to the boundary planes. 

We found interesting new behavior of the Casimir energy for GZ with PMC or PEC plates. 
We computed the Casimir energy for arbitrary \(\lambda\), both directly from the functional integral and from the energy-momentum tensor. 
Both for PMC and PEC plates, the Casimir force is attractive.
For the PEC case, we find that the Casimir energy is smaller (in absolute terms) than in the usual perturbative Yang-Mills case, and that \(\mathcal{E}_\text{Cas}^\text{GZ, PEC} \rightarrow \mathcal{E}_\text{Cas}^\text{YM}\) as \(\lambda\rightarrow0\), cfr.~\Cref{fig:energy-PEC}.
For the PMC case on the other hand, the Casimir energy is generally larger (in absolute terms) than in Yang-Mills, and \(\mathcal{E}_\text{Cas}^\text{GZ, PMC} \not \rightarrow \mathcal{E}_\text{Cas}^\text{YM}\) as \(\lambda\rightarrow0\), cfr.~\Cref{fig:energy-PMC}. 
This discontinuity with respect to taking the limit \(\lambda\rightarrow0\) is a vDVZ-like discontinuity, and can be traced back to the fact that the effective 3D boundary theory in GZ with PMC does not exhibit residual gauge freedom, in contrast to GZ with PEC and Yang-Mills. A similar effect is already present in the phenomenological Curci-Ferrari model for non-perturbative YM in the infrared, see \cite{Dudal2026}. We will come back to this issue in more detail in forthcoming work, including the dealing with (possible) boundary gauge copies~\cite{Dudal:2026uuv} and if this would give an extra $L$-dependent Gribov dynamics.

Another intriguing question arises as well. 
We have shown, in the current approximation, that the gap equation does not receive an \(L\)-dependent boundary term, and the main reason for this is that \(\ell_z\rightarrow\infty\), meaning that spacetime in the plate-direction is infinite. 
This leads us to wonder whether a dynamical dependence \(\lambda(L)\) might occur for a spacetime where \(\ell_z\) is finite, i.e.~a cylindrical spacetime where both plates are identified with each other. 
Besides this tantalizing possibility, studying this spacetime would require combining functional integral methods and computational techniques for periodic spacetime, best known from finite-temperature computations, an interesting endeavor in its own right.

We should also remark that in setting up the gap equation, we already made some (hidden) assumptions. In full generality, we reiterate that one should impose that all eigenvalues of the FP operator are positive. These eigenvalues are momentum dependent, thence the $\beta_x$ appearing in \eqref{heavi} would be functions, to be found from a functional minimization of the generalized version of \eqref{eq:threeGribov}. Needless to say, this would be prohibitively complicated, first of all because there is no closed expression for a generic eigenvalue, and second, it would require to compute with a gluon propagator depending on  generic momentum-dependent Gribov masses. 

We thus performed a perturbative analysis relative to the (presumably) most relevant zero momentum case. This is analogous to the original Gribov rationale \cite{Gribov78}: at small momentum, the coupling becomes large, so the ``potential term'' $\mathcal{M}_1$ in \eqref{fpsplit} can become strong enough to allow for ``bound states'' with negative ``energy''-eigenvalue. As argued, the zero momentum limit is a delicate one in the absence of translation invariance, so the small momenta should be kept until the very end. This means that, in principle, we could allow for generic $k_z$-dependent Gribov masses, with $k_z\to0$ understood. Even that looks very complicated, so a next best thing might be to introduce an educated guess for these $k_z$-dependent Gribov masses that allow for explicit computations. One could for example consider the Fourier transform of a Gribov mass that has different (constant) value between/outside the interfaces. Such approach goes well beyond the scope of this work, but it would certainly be interesting to pursue in future work.

When contemplating even further avenues of future research, an extension of the GZ framework comes to mind in which one takes into account the dynamical formation of dimension two condensates. This theory is known as refined Gribov-Zwanziger (RGZ)~\cite{Dudal05}, and leads to predictions in good agreement with the most recent lattice data~\cite{Dudal:2008sp,Dudal:2011gd}. It would be interesting to see if the features we found for the Casimir energy in GZ carry over to RGZ, or if refinement leads to a qualitatively different Casimir energy.


\section*{Acknowledgments}

The work of D.~Dudal was supported by KU Leuven IF project C14/21/087. The work of S.~Stouten was funded by FWO PhD-fellowship fundamental research (file number: 1132823N). P.~De~Fabritiis is grateful to S.~P.~Sorella for the discussions during the early stages of this work and acknowledges the National Council for Scientific and Technological Development – CNPq for the financial support (No. 402459/2024-5).


\appendix
\section{Most general gluon propagator} \label{appgluon}
In this appendix we compute the most general form the Landau-gauge gluon propagator can have in the presence of color rotation invariance and of rotation invariance and conservation of momentum in the directions perpendicular to the $z$-axis. Eventually, what we want an expression for is $k_\mu p_\nu\langle\tilde A_\mu^a(p-k)\tilde A_\nu^b(q-p)\rangle$.

Color rotation invariance dictates that this expectation value is equal to $\frac1{N^2-1} \delta^{ab} k_\mu p_\nu\langle\tilde A_\mu^c(p-k)\tilde A_\nu^c(q-p)\rangle$. Then, rotation invariance in the directions perpendicular to the $z$-axis gives that
\begin{equation}
	\begin{pmatrix} \langle\tilde A_i^c(p-k)\tilde A_j^c(q-p)\rangle & \langle\tilde A_i^c(p-k)\tilde A_z^c(q-p)\rangle \\ \langle\tilde A_z^c(p-k)\tilde A_j^c(q-p)\rangle & \langle\tilde A_z^c(p-k)\tilde A_z^c(q-p)\rangle \end{pmatrix} = \begin{pmatrix} \delta_{ij}A + (p-k)_i(p-k)_jB + \epsilon_{ijl}(p-k)_lE & (p-k)_iC \\ (p-k)_jD & \langle \tilde A_z^c(p-k)\tilde A_z^c(q-p)\rangle \end{pmatrix},\label{exchange}
\end{equation}
where $A$, $B$, $C$, and $D$ are unknown expressions involving gluon fields and momenta. We used that $\mathbf p-\mathbf k = \mathbf p-\mathbf q$ is the only three-dimensional vector available in the problem. Finally, it immediately follows from the trivial path integral subsitution $A_\mu^a(p)\to A_\mu^a(-p)$ that $\langle\tilde A_i^c(p-k)\tilde A_j^c(q-p)\rangle$ is even under exchange of the two gluon fields, meaning simultaneous exchange of $i\leftrightarrow j$, $\mathbf p-\mathbf k \leftrightarrow \mathbf q-\mathbf p = -\mathbf p-\mathbf k$, and $p_z-k_z\leftrightarrow q_z-p_z$. This means the $E$ term is allowed, but it will drop out when multiplying with $k_\mu p_\nu$. To compute the unknowns, take the trace of the three-dimensional part, also multiply with $p-k$ at the left, and also multiply with $q-p$ at the right. This gives five equations:
\begin{equation}
	\begin{cases}
		\langle\tilde A_i^c(p-k)\tilde A_i^c(q-p)\rangle = (d-1)A + (\mathbf p-\mathbf k)^2 B \\
		0 = (p-k)_jA + (\mathbf p-\mathbf k)^2(p-k)_jB + (p_z-k_z)(p-k)_jD \\
		0 = (\mathbf p-\mathbf k)^2C + (p_z-k_z)\langle \tilde A_z^c(p-k)\tilde A_z^c(q-p)\rangle \\
		0 = (p-k)_iA + (\mathbf p-\mathbf k)^2(p-k)_iB + (q_z-p_z)(p-k)_iC \\
		0 = -(\mathbf p-\mathbf k)^2D + (q_z-p_z)\langle \tilde A_z^c(p-k)\tilde A_z^c(q-p)\rangle
	\end{cases}.
\end{equation}
The zeros at the left-hand side appear due to the Landau gauge condition. These equations allow us to solve for all unknowns but $E$, which we do not need anyway. We finally find
\begin{multline} \label{perturb}
	k_\mu p_\nu\langle\tilde A_\mu^c(p-k)\tilde A_\nu^c(q-p)\rangle = \frac{\mathbf k^2}{d-2} \left(1-\frac{(\mathbf k\cdot(\mathbf p-\mathbf k))^2}{\mathbf k^2(\mathbf p-\mathbf k)^2}\right) \langle\tilde A_i^c(p-k)\tilde A_i^c(q-p)\rangle \\
	+ \left( \frac{\mathbf k^2}{d-2} \left(1-(d-1)\frac{(\mathbf k\cdot(\mathbf p-\mathbf k))^2}{\mathbf k^2(\mathbf p-\mathbf k)^2}\right) \frac{(p_z-k_z)(p_z-q_z)}{(\mathbf p-\mathbf k)^2} + \frac{\mathbf k\cdot(\mathbf p-\mathbf k) (2k_zq_z-p_zk_z-p_zq_z)}{(\mathbf p-\mathbf k)^2} + k_zq_z \right) \langle\tilde A_z^c(p-k)\tilde A_z^c(q-p)\rangle \;.
\end{multline}

\section{$K^{-1}$} \label{app:kinv}
We compute the operator $K^{-1}$ in \eqref{eq:K-inv}, but in the presence of the Gribov terms in \eqref{eq:threeGribov}. The quadratic part of the gluon action density is $\tfrac12\delta^{(3)}(\mathbf p-\mathbf q)\tilde A_\mu^a(p)\tilde A_\nu^a(-q)$ times
\begin{multline} \label{eq:kw}
	\delta(p_z-q_z) \begin{pmatrix} \left(p^2+\frac{\gamma_{\text{v}}^4}{p^2}\right)\delta_{ij}-\left(1-\tfrac1\alpha\right)p_ip_j & - \left(1-\tfrac1\alpha\right) p_i p_z \\ - \left(1-\tfrac1\alpha\right) p_z p_j & p^2-\left(1-\tfrac1\alpha\right)p_z^2 + \frac{\gamma_z^4}2 \left(\frac1{\mathbf p^2+2p_zk_z}+\frac1{\mathbf p^2-2p_zk_z}\right) \end{pmatrix} \\
	+ \frac{\gamma_{\text{br}}^4}2 \left(\frac{\delta(p_z-q_z+2k_z)}{\mathbf p^2+2p_zk_z}+\frac{\delta(p_z-q_z-2k_z)}{\mathbf p^2-2p_zk_z}\right) \begin{pmatrix} 0 & 0 \\ 0 & 1 \end{pmatrix} \;.
\end{multline}
In the above matrices, the first element (row or column) is for $\mu$ or $\nu$ not in the $z$ direction, and the second element is for $\mu$ or $\nu = z$. For later convenience, we averaged out the $zz$ elements over $\pm p$. We can invert the above matrix using the usual formulae for block matrix inversion, and we find in the limit $\alpha\to0$
\begin{equation} \label{eq:kwinv}
	\delta(p_z-q_z) \frac{p^2}{p^4+\gamma_{\text{v}}^4} \begin{pmatrix} \delta_{ij} - \frac{p_ip_j}{\mathbf p^2} & 0 \\ 0 & 0 \end{pmatrix} \\
	+ \frac1{\mathbf p^2\mathbf q^2} \mathcal K(\mathbf p,p_z,q_z) \begin{pmatrix} p_ip_z \\ -\mathbf p^2 \end{pmatrix} \begin{pmatrix} p_jq_z & -\mathbf p^2 \end{pmatrix} + \mathcal O(\alpha) \;,
\end{equation}
where the operator $\mathcal K$ is the operator inverse of
\begin{equation}
	\delta(p_z-q_z) \left(\frac{p^4}{\mathbf p^2} + \frac{\gamma_{\text{v}}^4p_z^2}{\mathbf p^2p^2} + \frac{\gamma_z^4}2 \left(\frac1{\mathbf p^2+(p_z+k_z)^2}+\frac1{\mathbf p^2+(p_z-k_z)^2}\right)\right) + \frac{\gamma_{\text{br}}^4}2 \left(\frac{\delta(p_z-q_z+2k_z)}{\mathbf p^2+2p_zk_z}+\frac{\delta(p_z-q_z-2k_z)}{\mathbf p^2-2p_zk_z}\right) + \mathcal O(\alpha) \;.
\end{equation}
Eventually we are interested in the limit $k_z\to0$, but due to the Dirac $\delta$'s, this limit is very nontrivial. For example, when computing $\frac d{d\beta} \det K$ for some $\beta$ or other, which is the kind of term that appears in the gap equation for the $\beta$s, we need to compute
\begin{equation}
	\int \frac{d^dp}{(2\pi)^d} \int \frac{d^dq}{(2\pi)^d} \frac{dK}{d\beta}(p,q) K^{-1}(q,p) \;.
\end{equation}
The operator $\mathcal K$ contains terms with $\delta(p_z-q_z+2nk_z)$ for all integers $n$, and the integral over $q_z$ will then give a sum over $n$. Putting $k_z=0$ too early would lead to the wrong result.


\bibliography{bibliography}


\end{document}